\documentclass[hyper,letterpaper,notoc]{JHEP3}

\usepackage{epsfig}
\usepackage{amsbsy}
\usepackage{varioref}
\usepackage{pifont}
\usepackage{amsmath}

\title{Multi-Gauge-Boson Vertices and Chiral Lagrangian Parameters
in Higgsless Models with Ideal Fermion Delocalization
}

\author{R. Sekhar Chivukula and Elizabeth H. Simmons\\
Department of Physics and Astronomy, Michigan State University\\
East Lansing, MI 48824, USA\\
	E-mail: \email{sekhar@msu.edu, esimmons@msu.edu}}

\author{Hong-Jian He\\
Department of Physics, University of Texas\\
Austin, TX 78712, USA\\
	E-mail: \email{hjhe@physics.utexas.edu}}

\author{Masafumi Kurachi and Masaharu Tanabashi\\
Department of Physics, Tohoku University\\
Sendai 980-8578, Japan\\
	E-mail: \email{kurachi@tuhep.phys.tohoku.ac.jp, tanabash@tuhep.phys.tohoku.ac.jp}}

\abstract{
Higgsless models with fermions whose $SU(2)$ properties are ``ideally delocalized,'' such
that the fermion's probability distribution is appropriately related to the $W$ boson wavefunction,
have been shown to minimize deviations in precision electroweak parameters. As contributions to the $S$ parameter vanish to leading order, current constraints
on these models arise from limits on deviations in multi-gauge-boson vertices.
We compute the form of the triple and quartic gauge
boson vertices in these models and show that these constraints provide lower 
bounds only of order a few hundred GeV on the masses of the lightest $KK$ resonances.  
Higgsless models
with ideal fermion delocalization provide an example of extended electroweak gauge
interactions with suppressed couplings of fermions to extra gauge-bosons, and these are the only models for which triple-gauge-vertex measurements provide meaningful constraints.
We relate the multi-gauge couplings to parameters of the electroweak chiral Lagrangian, and
the parameters obtained in these $SU(2) \times SU(2)$ models apply equally to the corresponding five dimensional gauge theory models of QCD. 
We also discuss the collider phenomenology of the $KK$ resonances in 
models with ideal delocalization. These resonances are 
found to be fermiophobic, therefore traditional direct collider searches are not sensitive to them and measurements of gauge-boson scattering will be needed to find them.  
}


\preprint{MSUHEP-050812\\
TU-750}

\begin{document}


\section{Introduction}
\label{sec:intro}

Higgsless models \cite{Csaki:2003dt} do just what their name suggests: they provide electroweak symmetry breaking, including unitarization of the scattering of longitudinal $W$ and $Z$ bosons, without employing a scalar Higgs \cite{Higgs:1964ia} boson.   In a class of well-studied models \cite{Agashe:2003zs,Csaki:2003zu} based on a five-dimensional
$SU(2) \times SU(2) \times U(1)$ gauge theory in a slice of Anti-deSitter space, 
electroweak symmetry breaking is encoded in the boundary conditions of the
gauge fields.  In addition to a massless photon and near-standard $W$ and $Z$ bosons, the spectrum includes an infinite tower of  additional massive vector bosons (the higher
Kaluza-Klein  or $KK$ excitations), whose exchange is responsible for unitarizing longitudinal $W$ and $Z$ boson scattering \cite{SekharChivukula:2001hz,Chivukula:2002ej,Chivukula:2003kq,He:2004zr}. 
The electroweak properties and collider phenomenology of many such models have been discussed in the literature \cite{Cacciapaglia:2004jz,Nomura:2003du,Barbieri:2003pr,Davoudiasl:2003me,Burdman:2003ya,Davoudiasl:2004pw,Barbieri:2004qk,Hewett:2004dv}.

An alternative approach to analyzing the properties of Higgsless models
\cite{Foadi:2003xa,Hirn:2004ze,Casalbuoni:2004id,Chivukula:2004pk,Perelstein:2004sc,Chivukula:2004af,Georgi:2004iy,SekharChivukula:2004mu}
is to use deconstruction 
\cite{Arkani-Hamed:2001ca,Hill:2000mu} and to 
compute the electroweak parameters $\alpha S$ and $\alpha T$  
\cite{Peskin:1992sw,Altarelli:1990zd,Altarelli:1991fk} in a 
related linear moose model \cite{Georgi:1985hf}. We have shown 
\cite{SekharChivukula:2004mu} how to compute 
all four of the leading zero-momentum electroweak parameters defined
by Barbieri et. al. \cite{Barbieri:2004qk} in a very general class of linear moose models.
Using these techniques, we showed \cite{SekharChivukula:2004mu} that a Higgsless model whose fermions are localized (i.e., derive their electroweak properties from a single site on the deconstructed lattice)  cannot simultaneously satisfy unitarity bounds and precision electroweak constraints unless the model includes extra light vector bosons with masses comparable to those of the $W$ or $Z$.

It has recently been proposed \cite{Cacciapaglia:2004rb,Foadi:2004ps,Cacciapaglia:2005pa}
 that the size of corrections to electroweak
processes may be reduced by including delocalized fermions. In deconstruction, a delocalized fermion is realized as a fermion whose $SU(2)$ properties arise from several
sites on the deconstructed lattice \cite{Chivukula:2005bn,Casalbuoni:2005rs}. 
We examined the case of a fermion whose $SU(2)$ properties arise from
two adjacent sites \cite{Chivukula:2005bn}, and confirmed that (even in that simple
case) it is possible to minimize the electroweak parameter $\alpha S$ by 
choosing a suitable amount of fermion delocalization.  In subsequent work \cite{SekharChivukula:2005xm},  we studied  the properties of deconstructed Higgsless models with fermions whose $SU(2)$ properties arise from delocalization over 
many sites of the deconstructed lattice.    In an arbitrary Higgsless model 
we showed that if the probability distribution
of the delocalized fermions 
is related to the $W$ wavefunction (a condition we call ``ideal'' delocalization), in flat space
and ${\rm AdS_5}$ for example (assuming the absence of brane kinetic energy terms for the gauge fields)
\begin{eqnarray}
\vert \psi(z)_{ideal}\vert^2 &\propto& \frac{1}{g_5^2}\chi_W(z)~,\ \ \ \ \ {\rm (flat\ space)}\nonumber \\
\vert \psi(z)_{ideal}\vert^2 &\propto& \frac{1}{z g_5^2}\chi_W(z)~,\ \ \ \ \ {({\rm AdS_5})}\nonumber
\end{eqnarray}
then deviations in precision electroweak parameters are minimized.\footnote{The vanishing
of $\alpha S$ to leading order for a ``flat'' fermion wavefunction in ${\rm AdS_5}$, which
is approximately related to the $W$ wavefunction in that model,  was noted also in
\cite{Agashe:2003zs,Cacciapaglia:2004rb}.} In particular, three ($\hat S$, $\hat T$, $W$) of the four leading zero-momentum precision electroweak parameters  defined by Barbieri, et. al. \cite{Barbieri:2004qk} vanish at tree-level.

This paper extends our analysis of ideal fermion delocalization in several ways.   We compute the form of the triple and quartic gauge boson vertices in these models and relate them to the parameters of the electroweak chiral Lagrangian.   As the symmetry structure of the models discussed is 
$SU(2) \times SU(2)$, the chiral Lagrangian parameters obtained apply equally to the corresponding
five dimensional gauge theory models of QCD \cite{Son:2003et,Hirn:2004ze,Chivukula:2004kg,Sakai:2004cn,Erlich:2005qh,DaRold:2005zs,Hirn:2005nr,Sakai:2005yt}. 
It is also
notable that Higgsless models with ideal delocalization
provide an example of a theory in which the chiral parameters
$\alpha_2$ (or $L_{9R}$) and $\alpha_3$ (or $L_{9L}$) are not equal.
We discuss the collider phenomenology of the $KK$ resonances in models with ideal delocalization in order to determine how experiments can constrain them in the absence of bounds from precision electroweak measurements.  

To begin, we provide additional details of the flat-space and warped-space
$SU(2)_A \times SU(2)_B$ models with brane kinetic terms that we will study.  We then discuss the wavefunctions, masses, and couplings of the $\gamma$, $W$ and $Z$ gauge bosons in the model; we also lay the groundwork for comparing the behavior of brane-localized and ideally-delocalized fermions.  We note that the $KK$ resonances lying above the $W$ and $Z$ bosons are fermiophobic.

In sections 4 and 5 we discuss the three-point and four-point gauge boson vertices which are crucial to the phenomenology of models with ideal fermion delocalization.  We first calcluate
the Hagiwara-Peccei-Zeppenfeld-Hikasa parameters \cite{Hagiwara:1986vm} to facilitate comparison with experiment.  We then cast our results into the language of the electroweak chiral lagrangian and find the values of the Longhitano  parameters \cite{Appelquist:1980vg,Longhitano:1980tm,Longhitano:1980iz,Appelquist:1980ix,Appelquist:1993ka}, and the corresponding parameters
as defined by Gasser and Leutwyler \cite{Gasser:1983yg}.  We also provide results for models based on 
a bulk $SU(2)$ gauge theory, and discuss how these are related to those for an 
$SU(2)_A \times SU(2)_B$
theory.

We then apply our results to analyze the unusual phenomenology of models with ideally-delocalized fermions. We show that current measurements of triple-gauge-boson vertices allow the lightest $KK$ resonances above the observed $W$ and $Z$ bosons to have masses of only a few hundred GeV.  Higgsless models
with ideal fermion delocalization provide an example of extended electroweak gauge
interactions with suppressed couplings of fermions to extra gauge-bosons, and these are the only models for which triple-gauge-vertex measurements provide meaningful constraints.\footnote{
Another interesting example is given in \cite{Georgi:2005dm}.}
Moreover,   
triple-gauge-vertex measurements are the only current source of constraints on Higgsless models with ideal fermion delocalization -- as mentioned earlier, these models are not constrained by precision electroweak tests.   Nor do existing direct searches for $W'$ and $Z'$ states constrain our models: such searches assume that the $W'$ and $Z'$ are produced and/or decay through their couplings to fermions -- but  the $KK$ resonances in models with ideal delocalization are fermiophobic.\footnote{The 
weakened coupling of fermions to $KK$ modes in the case of a ``flat" fermion wavefunction
in ${\rm AdS_5}$ is noted in \cite{Cacciapaglia:2004rb}.}  Studies of gauge-boson re-scattering \cite{Birkedal:2004au} will be needed in order to probe the $KK$ resonances of these models in more detail.
Our conclusions are summarized in section 7.  


\section{\protect{$SU(2)_A\times SU(2)_B$} Higgsless Models}
\label{sec:models}

\EPSFIGURE[t]{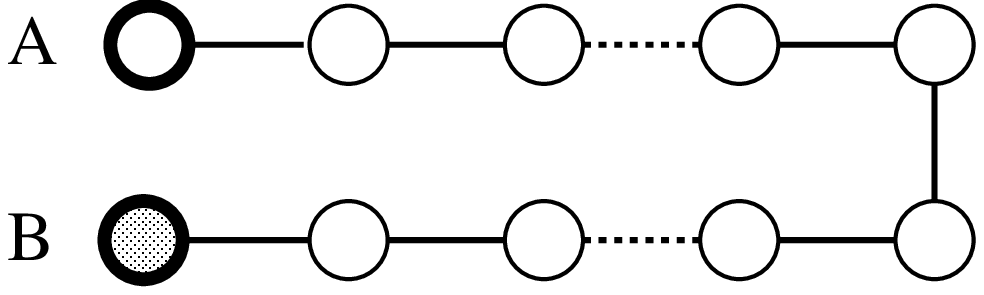,width=0.6\textwidth}
{Moose diagram \cite{Georgi:1985hf} for the deconstruction corresponding 
to the $SU(2)_A \times SU(2)_B$ models  analyzed in this paper. 
$SU(2)$ gauge groups are shown as open circles; $U(1)$ gauge group is shown as a shaded circle.
The brane kinetic energy terms are indicated by the thick circles. 
The fermions couple to a linear combination
of all of the $SU(2)$ groups, as well
as to the single $U(1)$ group. As drawn, this model does not have an interpretation
as a local higher-dimensional gauge theory;
however, the results of this model agree (up to corrections suppressed by
$M^4_W/M^4_{W_1}$) with a consistent higher dimensional theory  
based on an $SU(2) \times SU(2) \times U(1) $ gauge group in the bulk \cite{idealmodel}.
\label{fig:TheMoose}}

In this section we describe the five-dimensional $SU(2)_A \otimes SU(2)_B$
gauge theories, both in flat and warped spacetime,
which give rise to the Higgsless models discussed in this paper. The moose diagram 
\cite{Georgi:1985hf} for the deconstruction corresponding to these models is shown in
figure \ref{fig:TheMoose}. The fermions derive their $SU(2)_W$ properties from a linear
combination\footnote{More precisely, we associate each ordinary
fermion with the lightest (chiral) mode of a corresponding five-dimensional fermion field. 
Fermion delocalization
corresponds to a wavefunction of the lightest mode which extends into the bulk.}  of all of the $SU(2)$ groups -- in a manner related to the $W$ boson
profile as required for ideal delocalzation \cite{SekharChivukula:2005xm}. For the purposes
of the analyses presented here, we will take the $U(1)$ properties of the fermions to arise
from the single $U(1)$ group. As drawn, this model does not have an interpretation as
a local higher-dimensional gauge theory; 
however, the results of this model agree (up to corrections suppressed by
$M^4_W/M^4_{W_1}$) with a consistent higher dimensional theory  
based on an $SU(2) \times SU(2) \times  U(1)$ gauge group in the bulk \cite{idealmodel}.


\subsection{Flat}
\label{subsec:models:22:flat}

We begin by considering a five-dimensional $SU(2)_A \otimes
SU(2)_B$ gauge theory in flat space, in which the fifth dimension
(denoted by the coordinate $z$) is compactified on an interval
of length $\pi R$.
The action of the gauge theory is given by
\begin{eqnarray}
 S_{5D} &=& \int_{0}^{\pi R} dz \int d^4x 
 \left[\ 
 \frac{1}{g_{5A}^2} \left( -\frac{1}{4} 
		    W_{\mu\nu}^{Aa} W_{\alpha\beta}^{Aa} 
		    \eta^{\mu\alpha} \eta^{\nu\beta}+ \frac{1}{2}
		    W_{\mu z}^{Aa} W_{\nu z}^{Aa}\eta^{\mu\nu} 
		     \right)
\right.
 \nonumber\\ 
& & \hspace*{2.7cm} +
\left. \frac{1}{g_{5B}^2} \left( -\frac{1}{4} 
		    W_{\mu\nu}^{Ba} W_{\alpha\beta}^{Ba} 
		    \eta^{\mu\alpha}
			   \eta^{\nu\beta} + \frac{1}{2}
		    W_{\mu z}^{Ba} W_{\nu z}^{Ba}\eta^{\mu\nu}  \right)
\ \right]~,
\label{eq:lag}
\end{eqnarray}
where $a = 1, 2, 3$. The boundary conditions for $W_\mu^{A\, a}$ and
$W_\mu^{B\, a}$ are taken as 
\begin{equation}
\left. \partial_z W_\mu^{A\, a}(x,z) \right\vert_{z=0} = 0~, \ \ \ \ 
\left. W_\mu^{B\, 1,2}(x,z) \right\vert_{z=0} = 0~, \ \ \ \ 
\left. \partial_z W_\mu^{B\, 3}(x,z) \right\vert_{z=0} = 0~,
\label{eq:zboun}
\end{equation}
\begin{equation}
\left. \partial_z \left( \frac{1}{g_{5A}^2} W_\mu^{A\, a}(x,z) +
		   \frac{1}{g_{5B}^2}W_\mu^{B\, a}(x,z) 
       \right) \right\vert_{z=\pi R} = 0~, \ \ \ \ 
\left. \left( W_\mu^{A\, a}(x,z) - W_\mu^{B\, a}(x,z) \right)
\right\vert_{z=\pi R} = 0~.
\label{eq:rboun}
\end{equation}
The boundary conditions at $z=0$ explicitly break $SU(2)_A \otimes SU(2)_B$
down to $SU(2)_W \otimes U(1)_Y$, where we identify $SU(2)_W$ with $SU(2)_A$
and hypercharge with the $T^3$ component of $SU(2)_B$. The boundary conditions
at $z=\pi R$ break $SU(2)_A \otimes SU(2)_B$ to their diagonal subgroup; collectively
the boundary conditions leave only electromagentism unbroken.
We further introduce $SU(2)_A$ and 
$U(1)_Y$ kinetic term on the $z=0$ brane,
\begin{eqnarray}
 S\vert_{z=0} = \int_{0}^{\pi R} dz  
  \int d^4x \left[ -\,\delta(z-\epsilon)\frac{1}{4g_0^2}
			     W_{\mu\nu}^{A\, a}W_{\alpha\beta}^{A\,
			     a} \eta^{\mu\alpha}\eta^{\nu\beta}\right.
\nonumber \\
\left.
-\,\delta(z-\epsilon)\frac{1}{4g_Y^2}
			     W_{\mu\nu}^{B\, 3}W_{\alpha\beta}^{B\,
			     3}\eta^{\mu\alpha}\eta^{\nu\beta}
			     \right]~.  \ \ \ \ (\epsilon \rightarrow 0+)
\label{eq:brane_kinetic_term}
\end{eqnarray}
As the only $U(1)$ gauge symmetry exists on the $z=0$ brane,
the hypercharge of the fermions arises from couplings localized
at this brane (for a discussion of a consistent higher-dimensional gauge
theory allowing for ideal fermion delocalization, see \cite{idealmodel}).

The 5D fields $W_\mu^{A\, a}(x,z)$ and $W_\mu^{B\, a}(x,z)$ can be
decomposed into $KK$-modes,
\begin{eqnarray}
 W_\mu^{A\, 1,2}(x,z) &=& \sum_n W_\mu^{(n)\, 1,2}(x)\,
  \chi_{W_{(n)}}^{A}(z)~,   \nonumber\\  
 W_\mu^{B\, 1,2}(x,z) &=& \sum_n W_\mu^{(n)\, 1,2}(x)\,
  \chi_{W_{(n)}}^{B}(z)~, \nonumber \\ 
 W_\mu^{A\, 3}(x,z) &=& \gamma_\mu(x)\, \chi_{\gamma}^{A}(z) + \sum_n 
  Z_\mu^{(n)}(x)\, \chi_{Z_{(n)}}^{A}(z)~,   \nonumber\\ 
 W_\mu^{B\, 3}(x,z) &=& \gamma_\mu(x)\, \chi_{\gamma}^{B}(z) + \sum_n
  Z_\mu^{(n)}(x)\, \chi_{Z_{(n)}}^{B}(z)~.
\label{eq:mode_expansion}
\end{eqnarray}
Here $\gamma_\mu(x)$ is the photon, and $W_\mu^{(n)\, 1,2}(x)$ and 
$Z_\mu^{(n)}(x)$ are the $KK$ towers of the massive $W$ and $Z$ bosons,
the lowest of which correspond to the observed $W$ and $Z$ bosons. 
The mode functions $\chi_{W_{(n)}}^{A,B}(z)$, $\chi_{Z_{(n)}}^{A,B}(z)$
and $\chi_{\gamma}^{A,B}(z)$ obey the differential equations
\begin{eqnarray}
\frac{d^2}{dz^2}   \chi_{W_{(n)}}^{A,B}  + M_{W(n)}^2\, 
          \chi_{W_{(n)}}^{A,B}&=&  0~,
\label{eq:EOM_W}
\\ 
\frac{d^2}{dz^2} \chi_{Z_{(n)}}^{A,B}  + M_{Z(n)}^2\, 
          \chi_{Z_{(n)}}^{A,B} &=& 0~,
\label{eq:EOM_Z}
\\ 
\frac{d^2}{dz^2} \chi_{\gamma}^{A,B} &=& 0~,
\label{eq:EOM_photon}
\end{eqnarray}
and the boundary conditions 
\[
\left. \chi_{W_{(n)}}^{B}(z) \right\vert_{z=0} \ =\ 
\left. \partial_z \chi_{\gamma}^{B}(z) \right\vert_{z=0} \ =\  
\left. \partial_z \chi_{\gamma}^{A}(z) \right\vert_{z=0} \ =\ 0~,
\]
\begin{equation}
\left. \partial_z \chi_{W_{(n)}}^{A}(z) \right\vert_{z=0} \ =\ 
 - \frac{g_{5A}^2}{g_0^2} M_{W(n)}^2 \chi_{W_{(n)}}^{A}(0)~,
\label{eq:BC_TeV}
\end{equation}
\[
\left. \partial_z \chi_{Z_{(n)}}^{A}(z) \right\vert_{z=0} \ =\ 
 - \frac{g_{5A}^2}{g_0^2} M_{Z(n)}^2 \chi_{Z_{(n)}}^{A}(0)~,\]
\[
  \left. \partial_z \chi_{Z_{(n)}}^{B}(z) \right\vert_{z=0} \ =\ 
 - \frac{g_{5B}^2}{g_Y^2} M_{Z(n)}^2 \chi_{Z_{(n)}}^{B}(0)~,
\]
and
\begin{eqnarray}
\left. \partial_z \left( \frac{1}{g_{5A}^2} \chi_{W_{(n)}}^{A}(z) +
		   \frac{1}{g_{5B}^2} \chi_{W_{(n)}}^{B}(z)
       \right) \right\vert_{z=\pi R} &=& 0~,\ \ \ \ \ 
\left. \left( \chi_{W_{(n)}}^{A}(z) - \chi_{W_{(n)}}^{B}(z)
       \right) \right\vert_{z=\pi R} \ =\ 0~, \nonumber \\
\left. \partial_z \left( \frac{1}{g_{5A}^2} \chi_{Z_{(n)}}^{A}\,(z) +
		   \frac{1}{g_{5B}^2} \chi_{Z_{(n)}}^{B}\,(z)
       \right) \right\vert_{z=\pi R} &=& 0~, \ \ \ \ \ 
\left. \left( \chi_{Z_{(n)}}^{A}\,(z) - \chi_{Z_{(n)}}^{B}\,(z)
       \right) \right\vert_{z=\pi R} \ \, =\ 0~, \nonumber\\
\left. \partial_z \left( \frac{1}{g_{5A}^2} \chi_{\gamma}^{A}(z) +
		   \frac{1}{g_{5B}^2} \chi_{\gamma}^{B}(z)
       \right) \right\vert_{z=\pi R} &=& 0~,\ \ \ \ \ 
\left. \left( \chi_{\gamma}^{A}(z) - \chi_{\gamma}^{B}(z)
       \right) \right\vert_{z=\pi R} \ =\  0~.
\label{eq:BC_Planck}
\end{eqnarray}
Note that the boundary conditions at $z=0$ reflect the
presence of the $SU(2)_A \times U(1)_Y$ brane kinetic term Eq.\,(\ref{eq:brane_kinetic_term}).

Substituting  Eq.\,(\ref{eq:mode_expansion}) into $S_{5D}+S_{z=0}$ and
requiring that the 4D fields $W_\mu^{(n)}(x)$, $Z_\mu^{(n)}(x)$,
$\gamma_\mu(x)$ possess canonically normalized kinetic terms, we
see the mode functions are normalized as 
\begin{eqnarray}
 1 &=& \int_{0}^{\pi R}dz
  \left\{ \frac{1}{g_{5A}^2} \left|\chi_{W_{(n)}}^A(z)\right|^2 + 
          \frac{1}{g_{5B}^2} \left|\chi_{W_{(n)}}^B(z)\right|^2
 \right\}
 + \frac{1}{g_0^2} \left|\chi_{W_{(n)}}^A (0)\right|^2~,
\label{eq:norm_W}
\\
 1 &=& \int_{0}^{\pi R}dz 
  \left\{ \frac{1}{g_{5A}^2} \left|\chi_{\gamma}^A(z)\right|^2 \ \ \ + \  
          \frac{1}{g_{5B}^2} \left|\chi_{\gamma}^B(z)\right|^2 \ \ \,
 \right\}
 + \frac{1}{g_0^2} \left|\chi_{\gamma}^A (0)\right|^2
 + \frac{1}{g_Y^2} \left|\chi_{\gamma}^B (0)\right|^2~,
\label{eq:norm_photon}
\\
 1 &=& \int_{0}^{\pi R}dz
  \left\{ \frac{1}{g_{5A}^2} \left|\chi_{Z_{(n)}}^A(z)\right|^2 + 
          \frac{1}{g_{5B}^2} \left|\chi_{Z_{(n)}}^B(z)\right|^2 \ 
 \right\}
 + \frac{1}{g_0^2} \left|\chi_{Z_{(n)}}^A (0)\right|^2
 + \frac{1}{g_Y^2} \left|\chi_{Z_{(n)}}^B (0)\right|^2 .\ \ \ \ 
\label{eq:norm_Z}
\end{eqnarray}
Since the lightest massive $KK$-modes $(n=0)$ are identified as the observed $W$ and $Z$ bosons,
we will write
\begin{equation}
 M_W^2 \equiv M_{W_{(0)}}^2,\ \ \ \ \  M_Z^2 \equiv M_{Z_{(0)}}^2~,
\end{equation}
and
\begin{equation}
 \chi_W^{A,\,B} \equiv  \chi_{W_{(0)}}^{A,\,B}~,\ \ \ \ \  \chi_Z^{A,\,B} \equiv 
  \chi_{Z_{(0)}}^{A,\,B}~.
\end{equation}


\subsection{Warped}
\label{subsec:models:22:warp}

We will also consider a related model based on $SU(2)_A \otimes SU(2)_B$
in a truncated Anti-deSitter space.  We adopt the conformally flat metric
\begin{equation}
 ds^2 = \left( \frac{R}{z} \right)^2 \left( \eta_{\mu \nu} dx^\mu dx^\nu
				     - dz^2 \right)~,
\label{eq:adsmetric}
\end{equation}
for the ${\rm AdS_5}$ space, and require that the coordinate $z$ be 
restricted to the interval 
\begin{equation}
R \left(\equiv e^{-b/2} R'\right)\le z \le  R'~ .
\end{equation}
The endpoint $z=R$ is known as the ``Planck'' brane, while $z=R'$ is referred as
the ``TeV'' brane. We will assume a large hierarchy between $R$ and $R'$
($b \gg 1$). 

The 5D action of an $SU(2) \otimes SU(2)$ gauge
theory in this warped space is 
\begin{eqnarray}
 S_{5D} &=& \int_R^{R'} dz \frac{R}{z} \int d^4x 
 \left[\ 
 \frac{1}{g_{5A}^2} \left( -\frac{1}{4} 
		    W_{\mu\nu}^{Aa} W_{\alpha\beta}^{Aa} 
		    \eta^{\mu\alpha} \eta^{\nu\beta}+ \frac{1}{2}
		     W_{\mu z}^{Aa} W_{\nu z}^{Aa} \eta^{\mu\nu}\right)
\right.
 \nonumber\\ 
& & \hspace*{2.7cm} +
\left. \frac{1}{g_{5B}^2} \left( -\frac{1}{4} 
		    W_{\mu\nu}^{Ba} W_{\alpha\beta}^{Ba} \eta^{\mu\alpha}
			   \eta^{\nu\beta} + \frac{1}{2}
		    W_{\mu z}^{Ba} W_{\nu z}^{Ba}  \eta^{\mu\nu}\right)
\ \right] ~,
\label{eq:w:lag}
\end{eqnarray}
where $a = 1, 2, 3$. 
In order to arrange for non-trivial weak mixing angle, we further introduce a
$U(1)_Y$ kinetic term on the Planck brane ($z=R$),
\begin{equation}
 S_{Planck} = \int_R^{R'} dz\, \delta(z-R-\epsilon) 
  \int d^4x \,\frac{1}{g_Y^2} \left[ -\frac{1}{4}
			     W_{\mu\nu}^{B\, 3}W_{\alpha\beta}^{B\,
			     3}\eta^{\mu\alpha}\eta^{\nu\beta}
			     \right]~. \ \ \ \ (\epsilon \rightarrow 0+)
\label{eq:w:brane_kinetic_term}
\end{equation}
As before, the $U(1)_Y$ gauge symmetry exists only on the
left-hand (Planck) brane, and the hypercharge couplings of the fermions
therefore arise from couplings localized to that brane.

The 5D fields $W_\mu^{A\, a}(x,z)$ and $W_\mu^{B\, a}(x,z)$ can be
decomposed into $KK$-modes, exactly as before (see eqs. (\ref{eq:mode_expansion})).
However, the mode functions $\chi_{W_{(n)}}^{A,B}(z)$, $\chi_{Z_{(n)}}^{A,B}(z)$
and $\chi_{\gamma}^{A,B}(z)$ now obey modified  differential equations
\begin{eqnarray}
 0 &=& z\, \partial_z \left( \frac{1}{z}\, \partial_z 
          \chi_{W_{(n)}}^{A,B} \right) + M_{W(n)}^2\, 
          \chi_{W_{(n)}}^{A,B}~,
\label{eq:w:EOM_W}
\\ 
 0 &=& z\, \partial_z \left( \frac{1}{z}\, \partial_z 
          \chi_{Z_{(n)}}^{A,B} \right) + M_{Z(n)}^2\, 
          \chi_{Z_{(n)}}^{A,B} ~,
\label{eq:w:EOM_Z}
\\ 
 0 &=& z\, \partial_z \left( \frac{1}{z}\, \partial_z 
          \chi_{\gamma}^{A,B} \right)~.
\label{eq:w:EOM_photon}
\end{eqnarray}
The form of the boundary conditions at the TeV Brane, $z= R'$, is the same as the form of those at $z=\pi R$ in the flat case (eqns. (\ref{eq:BC_Planck})).  The boundary conditions at the Planck brane, $z=R$, are modified from those at $z=0$ in the flat case:
\[
  \left. \partial_z \chi_{W_{(n)}}^{A}(z) \right\vert_{z=R} \ =\ 
 \left. \partial_z \chi_{Z_{(n)}}^{A}(z) \right\vert_{z=R} \ =\ 
 \left. \partial_z \chi_{\gamma}^{A}(z) \right\vert_{z=R} \ =\ 0~,
\]
\begin{equation}
 \left. \chi_{W_{(n)}}^{B}(z) \right\vert_{z=R} \ =\ 
 \left. \partial_z \chi_{\gamma}^{B}(z) \right\vert_{z=R} \ =\ 0~,
\label{eq:w:BC_TeV}
\end{equation}
\[
  \left. \partial_z \chi_{Z_{(n)}}^{B}(z) \right\vert_{z=R} \ =\ 
 - \frac{g_{5B}^2}{g_Y^2} M_{Z(n)}^2 \chi_{Z_{(n)}}^{B}(R)~.
\]
Note that the boundary condition at $z=R$ reflects the
presence of the $U(1)_Y$ brane kinetic term Eq.\,(\ref{eq:w:brane_kinetic_term}).

Substituting Eq.\,(\ref{eq:mode_expansion}) into $S_{5D}+S_{Planck}$ and
requiring that the 4D fields $W_\mu^{(n)}(x)$, $Z_\mu^{(n)}(x)$, and
$\gamma_\mu(x)$ possess canonically normalized kinetic terms, we
see the mode functions are normalized as 
\begin{eqnarray}
 1 &=& \int_{R}^{R'}dz \left(\frac{R}{z}\right)  
  \left\{ \frac{1}{g_{5A}^2} \left|\chi_{W_{(n)}}^A(z)\right|^2 + 
          \frac{1}{g_{5B}^2} \left|\chi_{W_{(n)}}^B(z)\right|^2
 \right\}~,
\label{eq:w:norm_W}
\\
 1 &=& \int_{R}^{R'}dz \left(\frac{R}{z}\right)  
  \left\{ \frac{1}{g_{5A}^2} \left|\chi_{\gamma}^A(z)\right|^2 \ \ \ + \  
          \frac{1}{g_{5B}^2} \left|\chi_{\gamma}^B(z)\right|^2 \ \ \,
 \right\}
 + \frac{1}{g_Y^2} \left|\chi_{\gamma}^B (z=R)\right|^2~,
\label{eq:w:norm_photon}
\\
 1 &=& \int_{R}^{R'}dz \left(\frac{R}{z}\right)  
  \left\{ \frac{1}{g_{5A}^2} \left|\chi_{Z_{(n)}}^A(z)\right|^2 + 
          \frac{1}{g_{5B}^2} \left|\chi_{Z_{(n)}}^B(z)\right|^2 \ 
 \right\}
 + \frac{1}{g_Y^2} \left|\chi_{Z_{(n)}}^B (z=R)\right|^2~.
\label{eq:w:norm_Z}
\end{eqnarray}
The lightest massive $KK$-modes $(n=0)$ are, again, identified as the observed $W$ and $Z$ bosons. 


\section{Couplings, Masses, and Wavefunctions}

In this section we derive expressions for the couplings, masses, and
wavefunctions of the photon, $W$ and $Z$-bosons for the Higgsless Models
described above.

\subsection{The photon and the bare weak angles}
\label{subsec:mass:photon} 

Let us start with the zero-mode $\chi_\gamma^{A,\,B}$, which is
identified as the photon.
The solution of the differential equation Eq.~(\ref{eq:EOM_photon}) 
obeying boundary conditions Eq.~(\ref{eq:BC_Planck}) and 
Eq.~(\ref{eq:BC_TeV}) in the flat case is
\begin{equation}
 \chi_\gamma^A(z) =  \chi_\gamma^B(z) = {\rm constant}~.
\end{equation}
The identical result is obtained in the warped case.

From the normalization conditions in Eq.~(\ref{eq:norm_photon}) and (\ref{eq:w:norm_photon}), we find 
\begin{equation} 
 \chi_\gamma^A(z) =  \chi_\gamma^B(z) = 
\left[\pi R\left(\frac{1}{{g}_{5A}^2}+\frac{1}{{g}_{5B}^2}\right)\ 
 + \frac{1}{g_0^2}+ \frac{1}{g_Y^2}\right]^{-\frac{1}{2}}~,\ \ \ \ {\rm (flat)}
\label{eq:photon_wavefunction}
\end{equation}
\begin{equation}
 \chi_\gamma^A(z) =  \chi_\gamma^B(z) = 
  \left[\frac{bR}{2}\left(\frac{1}{g_{5A}^2}+\frac{1}{g_{5B}^2}\right)
  + \frac{1}{g_Y^2}\right]^{-\frac{1}{2}}~.\ \ \ \ ({\rm AdS_5})
\label{eq:w:photon_wavefunction}
\end{equation}
Since $\chi_\gamma^{A,\,B}(z)$ is the mode function of photon, these
expressions  yield the electromagnetic coupling $e$:
\begin{equation}
 \frac{1}{e^2} =
\pi R \left(\frac{1}{{g}_{5A}^2}+\frac{1}{{g}_{5B}^2}\right)\
  + \frac{1}{g_0^2}+ \frac{1}{g_Y^2}~,\ \ \ \ {\rm (flat)}
\end{equation}
\begin{equation}
 \frac{1}{e^2} =
  \frac{bR}{2}\left(\frac{1}{g_{5A}^2}+\frac{1}{g_{5B}^2}\right) +
  \frac{1}{g_Y^2} ~.\ \ \ \ ({\rm AdS_5})
\end{equation}
We can immediately identify quantities $c_0$, $s_0$ that satisfy the relation
\begin{equation}
 s_0^2 + c_0^2 = 1~,
\end{equation}
and may be interpreted as the cosine and sine of the bare weak mixing angle:
\begin{equation}
 c_0^2 =  \frac{g_0^2}{g_0^2+g_Y^2}~,\ \ \ \ 
 s_0^2 =  \frac{g_Y^2}{g_0^2+g_Y^2}~, \ \ \ \ {\rm (flat)}
\end{equation}
\begin{equation}
 \frac{c_0^2}{e^2} = \left(\frac{bR}{2}\right) \frac{1}{g_{5B}^2} 
  + \frac{1}{g_Y^2}~,\ \ \ \ 
 \frac{s_0^2}{e^2} = \left(\frac{bR}{2}\right) \frac{1}{g_{5A}^2}~.\ \ \ \ ({\rm AdS_5})
\end{equation}

The ratio of the left-handed and right-handed squared-couplings will be useful in
our later analysis
\begin{equation}
 \kappa \equiv \frac{g_{5B}^2}{g_{5A}^2}~, \ \ \ \ \ \ 
\end{equation}
of both the flat and warped cases.   For the flat-space model, we will
also find it convenient to define the dimensionless quantities :
\begin{equation}
\tilde{g}_{5A}^2 \equiv \frac{g_{5A}^2}{\pi R}~,\ \ \ \ \ \ \ 
\tilde{g}_{5B}^2 \equiv \frac{g_{5B}^2}{\pi R}~,
\end{equation}
\begin{equation}
 {\tilde M_W} \equiv \pi R M_W ~, \ \ \ \ \ \ 
 {\tilde M_Z} \equiv \pi R M_Z~, \ \ \ \ \ \ 
 \tilde{z} \equiv \frac{z}{\pi R}~,
 \label{eq:fmwi}
\end{equation}
and the ratio of couplings:
\begin{equation}
 \lambda \equiv \frac{g_0^2}{\tilde{g}_{5A}^2+\tilde{g}_{5B}^2}~.
\label{eq:lambda}
\end{equation}

\subsection{The W and Z}
\label{subsec:mass:wz}

From the differential equations~(\ref{eq:EOM_W}, \ref{eq:EOM_Z}), 
boundary conditions~(\ref{eq:BC_Planck}, \ref{eq:BC_TeV}), 
and normalization conditions~(\ref{eq:norm_W}, \ref{eq:norm_Z}),
we can determine the masses and wavefunctions of the $W$ and $Z$
bosons in the flat-space model :

\begin{equation}
 {\tilde M_W}^2 = 
   \lambda - \left(\frac{1+3\kappa}{3}\right)\lambda^2 
 \ +\ O(\lambda^3) ,\
 \label{eq:fmwii}
\end{equation}

\begin{eqnarray}
 \chi_W^A(z) &=& C_{W}
    \left[ 
       1-\frac{1}{(1+\kappa)}{\tilde z} \right. \nonumber\\
       & & \left. \ \ \ \ \ 
       + {\tilde M_W}^2 
            \left\{ 
              \frac{1}{3}\left(\frac{1+3\kappa}{1+\kappa}\right){\tilde z}
              -\frac{1}{2}{\tilde z}^2 
              +\frac{1}{6(1+\kappa)}{\tilde z}^3
            \right\}
       + O({\tilde M_W}^4)
    \right]~,
\label{eq:WL}
\\
 \chi_W^B(z) &=& C_{W} \left(\frac{\kappa}{1+\kappa}\right)
    \left[ 
         {\tilde z} 
       + {\tilde M_W}^2 
            \left\{ 
              \frac{2}{3}{\tilde z}
              -\frac{1}{6}{\tilde z}^3
            \right\}
       + O({\tilde M_W}^4)
    \right]~,
\label{eq:WR}
\end{eqnarray}

\begin{equation}
  C_{W} = g_0
    \left[ 
        1-\frac{1+3\kappa}{6}{\tilde M_W}^2 + O({\tilde M_W}^4)
    \right]~,
\end{equation}

\begin{equation}
 {\tilde M_Z}^2 =
 \frac{\lambda}{c_0^2} - 
 \left(\frac{\lambda}{c^2}\right)^2
   \frac{s^4(3+\kappa)-4s^2c^2\kappa+c^4(\kappa+3\kappa^2)}
        {3\kappa}
 \ +\ O(\lambda^3) ,
\end{equation}

\begin{eqnarray}
 \chi_Z^A(z) &=& C_{Z}
    \left[ 
       1-\frac{1}{c_0^2(1+\kappa)} {\tilde z} 
       + {\tilde M_Z}^2 
            \left\{ 
                \frac{s^4(3+\kappa)-4s^2c^2\kappa+c^4(\kappa+3\kappa^2)}
                     {3c^2\kappa(1+\kappa)} {\tilde z} 
            \right.
    \right.
    \nonumber\\ & & 
    \left.
            \left. \hspace{5cm}
                  -\frac{1}{2}{\tilde z}^2 
              +\frac{1}{6c^2(1+\kappa)}{\tilde z}^3
            \right\}
       + O({\tilde M_Z}^4)\ 
    \right]~,
\\
 \chi_Z^B(z) &=& -C_{Z}\frac{s_0^2}{c_0^2} 
   \left[ 
       1-{\tilde M_Z}^2\ \frac{s^4(1+\kappa)+s^2c^2(1-\kappa^2)
                              -c^4(\kappa+\kappa^2)}{\kappa}
   \right] \nonumber\\
  & & 
   \times \left[ 
       1-\frac{\kappa}{s_0^2(1+\kappa)}{\tilde z} 
       + {\tilde M_Z}^2 
            \left\{ 
                \frac{s^4(3+\kappa)-4s^2c^2\kappa+c^4(\kappa+3\kappa^2)}
                     {3s^2(1+\kappa)} {\tilde z} 
            \right.
    \right.
    \nonumber\\ & & 
    \left.
            \left. \hspace{5cm}
                  -\frac{1}{2}{\tilde z}^2 
              +\frac{\kappa}{6s^2(1+\kappa)}{\tilde z}^3
            \right\}
       + O({\tilde M_Z}^4)\ 
    \right]~,
\end{eqnarray}

\begin{equation}
  C_{Z} = g_0 c_0
    \left[ 
        1+{\tilde M_Z}^2\ 
        \frac{s^4(3+5\kappa)-s^2c^2(2\kappa+6\kappa^2)
              -c^4(\kappa+3\kappa^2)}{6\kappa}          
      + O({\tilde M_Z}^4)
    \right]~.
\end{equation}

Employing the analogous equations (\ref{eq:w:EOM_W}-\ref{eq:w:norm_Z})
for the warped-space model yields: 

\begin{equation}
 \left(M_W R'\right)^2 = \left(\frac{4}{b}\right)\frac{1}{(1+\kappa)} 
  \left\{1+\frac{3}{2b}\frac{1}{(1+\kappa)}\right\} \ +\ 
  O\left(\frac{1}{b^3}\right) ~,
\end{equation}

\begin{eqnarray}
 \chi_W^A &=& \frac{e}{s_0}
  \left\{1+\frac{3}{8}\frac{1}{(1+\kappa)}\left(\frac{2}{b}\right)\right\} 
  \nonumber\\
&\times&
 \left[1-\frac{1}{2}\left(\ln{\frac{z}{R}}-\frac{1}{2}\right)
  \left(M_W z\right)^2    + 
  \frac{1}{16}\left(\ln{\frac{z}{R}}-\frac{5}{4}\right)
  \left(M_W z\right)^4 \ + \cdots\ \ 
\right],\label{eq:W_A}\\
& &\nonumber\\
& &\nonumber\\
 \chi_W^B &=& \frac{e}{s_0} \kappa 
  \left\{1+\frac{3}{8}\frac{1}{(1+\kappa)}\left(\frac{2}{b}\right)\right\} 
  \left[\frac{1}{2}\left(M_W z\right)^2\left(\frac{b}{2}\right)
	  -\frac{1}{16}\left(M_W z\right)^4\left(\frac{b}{2}\right) \ +
	  \cdots\ \   \right]~,\label{eq:W_B}\\ 
& &\nonumber
\end{eqnarray}

\begin{equation}
 \left(M_Z R'\right)^2 =
  \frac{1}{c_0^2}\left(\frac{4}{b}\right)\frac{1}{(1+\kappa)}  
  \left\{1+\frac{3}{2b}\frac{1}{c_0^2}\frac{1}{(1+\kappa)}\right\} \ +\ 
  O\left(\frac{1}{b^3}\right) ~,
\end{equation}

\begin{eqnarray}
 \chi_Z^A &=& e\frac{c_0}{s_0}
  \left\{1+\frac{3}{8c_0^2}\frac{1}{(1+\kappa)}\left(\frac{2}{b}\right)\right\}
  \nonumber\\
&\times&
 \left[1-\frac{1}{2}\left(\ln{\frac{z}{R}}-\frac{1}{2}\right)
  \left(M_Z z\right)^2    + 
  \frac{1}{16}\left(\ln{\frac{z}{R}}-\frac{5}{4}\right)
  \left(M_Z z\right)^4 \ + \cdots\ \ 
\right],\\
& &\nonumber\\
& &\nonumber\\
 \chi_Z^B &=& e\frac{s_0}{c_0} \left\{1+\frac{3}{8c_0^2}\frac{1}{(1+\kappa)}\left(\frac{2}{b}\right)\right\}
   \left[-1+\frac{1}{2}\left(\ln{\frac{z}{R}}-\frac{1}{2} +
			\frac{e^2}{s_0^2}\frac{\kappa}{g_Y^2}\left(\frac{b}{2}\right)\right)    
  \left(M_Z z\right)^2 \right. \nonumber
\\ 
  & &\hspace{3.5cm}  \left. -
  \frac{1}{16}\left(\ln{\frac{z}{R}}-\frac{5}{4} + \frac{e^2}{s_0^2}\frac{\kappa}{g_Y^2}\left(\frac{b}{2}\right)
	      \right)
  \left(M_Z z\right)^4 \ + \cdots\ \ \right] ~.
\end{eqnarray}

\subsection{Fermion profiles}
\label{subsec:mass:fermion}

It will also be necessary to introduce the wavefunction assumed for the left-handed components
of the ordinary fermions (the right-handed components are assumed to
be localized on the same brane as the localized hypercharge kinetic
energy term).  We explore two possibilities, brane-localized fermions (which are phenomenologically
disfavored \cite{SekharChivukula:2004mu} because this predicts sizeable precision electroweak corrections) and 
ideally-delocalized fermions as defined in \cite{SekharChivukula:2005xm} (which are favored because they yield
small electroweak corrections).

The profile of the brane-localized fermion is expressed as (note that we  include any 
relevant metric factors in our definition of $|\psi|^2$ -- see Appendix \ref{appendix1})
\begin{eqnarray}
  \left| \psi^A_{\rm localized}(z)\right|^2 &=& \delta(z) ,\ \ \ \ \  {\rm (flat)}\\
  \left| \psi^A_{\rm localized}(z)\right|^2 &=& \delta(z-R) ,\ \ \ \ \  ({\rm AdS_5})\\
  \left| \psi^B_{\rm localized}(z)\right|^2 &=& 0~. \ \ \ \ \ {\rm (flat)\ \  or}\ \  ({\rm AdS_5})
\end{eqnarray}

In contrast, the wavefunction of an ideally delocalized fermion is specifically related to the 
$W$ wavefunction.  For a flat-space model, the relationship is:
\begin{eqnarray}
  \left| \psi^A_{\rm ideal}(z)\right|^2 &=& C_{\rm ideal} \left( \frac{1}{g_0^2}\delta(z) + \frac{1}{g_{5A}^2} \right) \chi_W^A(z) ,\\
  \left| \psi^B_{\rm ideal}(z)\right|^2 &=& C_{\rm ideal} \ \frac{1}{g_{5B}^2} \chi_W^B(z)~,
\end{eqnarray}
while in an $AdS_5$ model, it is 
\begin{eqnarray}
  \left| \psi^A_{\rm ideal}(z)\right|^2 &=& C_{\rm ideal}
   \left(\frac{R}{z}\right) \frac{1}{g_{5A}^2} \chi_W^A(z) ,\\
  \left| \psi^B_{\rm ideal}(z)\right|^2 &=& C_{\rm ideal} 
   \left(\frac{R}{z}\right) \frac{1}{g_{5B}^2} \chi_W^B(z)~.
\end{eqnarray}
The constant, $C_{\rm ideal}$  is determined from the fermion normalization 
condition (see Appendix \ref{appendix1})
\begin{equation}
  \int_0^{\pi R}dz \left\{ \left| \psi^A_{\rm ideal}(z)\right|^2 + \left| \psi^B_{\rm ideal}(z)\right|^2\right\} = 1,
\end{equation}
to be 
\begin{equation}
 C_{\rm ideal} = g_0 \left[ 1 - \frac{\lambda}{6}(5+3\kappa) \right]~, \ \ \ \ \ {\rm (flat)}
\end{equation}
\begin{equation}
 C_{\rm ideal} = \frac{e}{s_0} \left\{1-\frac{3}{8}\frac{1}{(1+\kappa)}
				\left(\frac{2}{b}\right)\right\} . \ \ \ \ \ ({\rm AdS_5})
\end{equation}

\subsection{\protect{$G_F$, $s_z^2$, and $c_z^2$}}
\label{subsec:mass:gf}

The coupling between the $W$ boson and  fermion $\psi$ is measured by the overlap integral
\begin{equation}
 g_W =\int_a^{b}dz \left\{
\left| \psi^A(z)\right|^2 \chi_W^A(z) + \left| \psi^B(z)\right|^2 \chi_W^B(z) 
\right\} ,\\
\end{equation}
where the limits of integration $[a,b]$ are $[0,\ \pi R]$ for flat space models and $[R,\ R']$ for warped space models.   Given the forms of the fermion and $W$-boson wavefunctions above, one finds 
directly that the couplings are as follows:
\begin{eqnarray}
g_W &=&  \chi^A_W(z=a)~,\ \ \ \ \ {\rm (brane\ localized)} \\
g_W &=& C_{ideal}~, \ \ \ \ \ {\rm (ideal)}
\end{eqnarray}
where $a$ is the z-coordinate of the brane on which the fermion is localized ($0$ for flat-space; $R$ for $AdS_5$) and $C_{ideal}$ is the fermion normalization constant from section \ref{subsec:mass:fermion}.

The contributions to $G_F$ arising from the exchange of higher-mass $KK$ modes are
suppressed in these models (the $KK$ wavefunctions are suppressed precisely where the
fermion wavefunctions are concentrated), and therefore, we can express the Fermi constant to
this order as 
\begin{equation}
 4\sqrt{2} G_F =\frac{g_W^2}{M_W^2} .
 \end{equation}
 Drawing on the values of $g_W$ and $M_W$ from our earlier discussion, we deduce that
 \begin{eqnarray}
  4\sqrt{2} G_F &=&  \frac{g_0^2}{\lambda}(\pi R)^2 \left\{1 + O\left(\lambda^2\right)\right\} , \ \ \ \ {\rm (flat, brane\ localized)}\\
  4\sqrt{2} G_F  &=& \frac{g_0^2}{\lambda}(\pi R)^2  \left[1 - \frac{4}{3} \lambda\right]~,
\ \ \ {\rm (flat, ideal)}\\
  4\sqrt{2} G_F&=& \frac{e^2}{s_0^2} (R')^2\frac{b}{4}(1+\kappa)\left\{1 +
 O\left(\frac{1}{b^2}\right)\right\}~, \ \ \ {\rm (AdS_5, brane\ localized)}\\
   4\sqrt{2} G_F&=& \frac{e^2}{s_0^2} (R')^2\frac{b}{4}(1+\kappa) 
\left\{1-\frac{3}{2}\frac{1}{(1+\kappa)}\left(\frac{2}{b}\right)\right\}~. \ \ \ {\rm (AdS_5, ideal)}
 \end{eqnarray}
 Note that in the expressions for brane localized fermions, the leading order correction vanishes; this is the $\lambda^1$ term for the flat space model and the $b^{-1}$ term for the warped space model.

The standard definition of the weak mixing angle is given by
\begin{equation}
 s_Z^2 c_Z^2 = \frac{e^2}{4\sqrt{2}G_F M_Z^2}.
\end{equation}
The deviation of $s_Z^2$ from its bare value $s_0^2$ is parametrized by 
\begin{equation}
 s_Z^2 = s_0^2 + \Delta ,\ \ \ \ c_Z^2 = c_0^2 - \Delta.
\end{equation}
Since we already have calculated $M_Z$ and $G_F$, we are able to
evaluate $\Delta$ for each model directly:
\begin{eqnarray}
 \Delta &=& \frac{\lambda}{3} \left(\frac{s^2}{c^2-s^2}\right)
  \left\{3c^2(c^2-s^2)\kappa + (c^4-10c^2s^2+s^4) -
   3(c^2-s^2)s^2\frac{1}{\kappa}\right\},\label{eq:Delta_localized}
\\
  & &\hspace{8cm}{\rm (flat,\ brane\ localized)}\nonumber\\
 \Delta &=& \frac{\lambda}{3} \left(\frac{s^2}{c^2-s^2}\right)
  \left\{3c^2(c^2-s^2)\kappa + (5c^4-6c^2s^2+s^4) -
   3(c^2-s^2)s^2\frac{1}{\kappa}\right\}.\label{eq:Delta_ideal}
\\
  & &\hspace{8cm}{\rm (flat,\ ideal)}\nonumber\\
\Delta &=& -\frac{3}{4}\frac{s_0^2}{c_0^2-s_0^2}\frac{1}{(1+\kappa)}
  \left(\frac{2}{b}\right) \ \ \ \ \ {\rm (AdS_5,\ brane\ localized)}\\
\Delta  &=& \frac{3}{4}s_0^2\frac{1}{(1+\kappa)}
  \left(\frac{2}{b}\right) \ \ \ \ \ {\rm (AdS_5,\ ideal)}.
  \end{eqnarray}

\subsection{\protect{$\alpha S$}}
\label{subsec:mass:as}

Ideal fermion delocalization takes its name from the fact that it is constructed in such a way as to minimize the size of precision electroweak corrections, as discussed in \cite{SekharChivukula:2005xm}.  For example, the $S$ parameter can be calculated  from the $T_3$-$Y$ correlation function 
\cite{Chivukula:2004af,SekharChivukula:2004mu} arising from
$Z$-exchange, and we therefore find:
\begin{equation}
 1+\frac{\alpha S}{4 s_Z^2 c_Z^2} = -\frac{1}{e^2}\ \chi_Z^B(z=a)\ 
  \int_a^{b}dz \left\{ 
 \left| \psi^A(z) \right|^2 \chi_Z^A(z)  +  \left| \psi^B(z) \right|^2 \chi_Z^B(z) \right\}.
\end{equation}
where the limits of integration $[a,b]$ are $[0,\ \pi R]$ for flat space and $[R,\ R']$ for warped space.
Using the forms of the wavefunctions from sections \ref{subsec:mass:wz} and \ref{subsec:mass:fermion},
we calculate:
\begin{eqnarray}
\alpha S &=& \frac{8}{3}\ \lambda\ s^2~, \ \ \ \ \ {\rm (flat,\ brane\ localized)}\\
\alpha S &=& 0 \ +\ O\left(\lambda^2\right)~, \ \ \ \ \ {\rm (flat,\ ideal)}\\
\alpha S &=& 3 s^2 \frac{1}{(1+\kappa)}\left(\frac{2}{b}\right)~, \ \ \ \ \ {\rm (AdS_5,\ brane\ localized)}\\
\alpha S &=&  0 \ +\ O\left(\frac{1}{b^2}\right)~. \ \ \ {\rm (AdS_5,\ ideal)}
\end{eqnarray}
Note that for ideal delocalization, $\alpha S$ vanishes to leading order.


\section{Multi-Gauge-Boson Vertices}
\label{sec:results}

\subsection{Triple-Gauge Boson Interactions: Notation}
\label{subsec:vertices:lang}

To leading order, in the absence of CP-violation, the triple gauge boson vertices
may be written in the Hagiwara-Peccei-Zeppenfeld-Hikasa triple-gauge-vertex 
notation \cite{Hagiwara:1986vm}
\begin{eqnarray}
{\cal L}_{TGV} & = & -ie\frac{c_Z }{s_Z}\left[1+\Delta\kappa_Z\right] W^+_\mu W^-_\nu Z^{\mu\nu}
- ie \left[1+\Delta \kappa_\gamma\right] W^+_\mu W^-_\nu A^{\mu\nu} \cr
&-& i e \frac{c_Z}{s_Z} \left[ 1+\Delta g^Z_1\right](W^{+\mu\nu}W^-_\mu - W^{-\mu\nu} W^+_\mu)Z_\nu 
\label{eq:tgvlag} \\
&-& ie (W^{+\mu\nu}W^-_\mu - W^{-\mu\nu} W^+_\mu) A_\nu~, \nonumber
\end{eqnarray}
where the two-index tensors denote the Lorentz field-strength
tensor of the corresponding field. In the standard model, 
$\Delta\kappa_Z = \Delta\kappa_\gamma = \Delta g^Z_1 \equiv 0$.

In any vector-resonance model, such as the Higgsless models considered here,
the interactions (\ref{eq:tgvlag}) come from re-expressing the nonabelian couplings in the kinetic
energy terms in the original Lagrangian ({\it e.g.}, eqns. (\ref{eq:lag}) and 
(\ref{eq:brane_kinetic_term}), or (\ref{eq:w:lag}) and (\ref{eq:w:brane_kinetic_term}))
in terms of the mass-eignestate fields. In this case one obtains equal contributions
to the deviations of the first and third terms, and the second and fourth terms
in eqn. (\ref{eq:tgvlag}). In addition the coefficient of the fourth term is fixed by electromagnetic
gauge-invariance, and therefore in these models we find
\begin{equation}
\Delta \kappa_\gamma \equiv 0 \ \ \ \  \ \Delta\kappa_Z \equiv \Delta g^Z_1~.
\label{eq:tgv_relations}
\end{equation}
%

\subsection{Triple gauge boson vertices}
\label{subsec:vertices:22:flat}

We are now ready to calculate the triple gauge boson vertices.  In the flat-space
case, the relevant integral (using $V$ to stand for either $\gamma$ or $Z$) is:
\begin{equation}
 g_{V WW} = \int_0^{\pi R}dz
  \left\{ \frac{1}{g_{5A}^2} \chi_{V}^A(z)
   \left|\chi_{W}^A(z)\right|^2 +  
          \frac{1}{g_{5B}^2} \chi_{V}^B(z)
	  \left|\chi_{W}^B(z)\right|^2 
 \right\} 
+ \frac{1}{g_{0}^2} \chi_{V}^A(0)
   \left|\chi_{W}^A(0)\right|^2
   \end{equation}
while in the warped-space case, one writes instead:
\begin{equation}
g_{V WW} = \int_R^{R'}dz \left(\frac{R}{z}\right)
  \left\{ \frac{1}{g_{5A}^2} \chi_{V}^A(z)
   \left|\chi_{W}^A(z)\right|^2 +  
          \frac{1}{g_{5B}^2} \chi_{V}^B(z)
	  \left|\chi_{W}^B(z)\right|^2 
 \right\} \ .
 \end{equation}

By inserting the forms of the  gauge boson wavefunctions from sections \ref{subsec:mass:photon} and  \ref{subsec:mass:wz}, we verify that
\begin{equation}
 g_{\gamma WW} = e .
\end{equation}
in all cases, and find that the $ZWW$ coupling has the forms
\begin{eqnarray}
 g_{ZWW} &=&
  e\frac{c_0}{s_0}\left[ 1 + 
  \lambda\ \frac{2+5\kappa+\kappa^2 - 2 c^2 (1+\kappa)^3}{4 c^2
  \ \kappa (1+\kappa)} \right] \ \ \ \ \ {\rm (flat)} \label{eq:ZWW}\\
  g_{ZWW} &=&
  e\frac{c_0}{s_0}\left\{1+\frac{5}{24}\frac{1}{c_0^2}
		   \frac{(1-\kappa)}{(1+\kappa)^2}
		   \left(\frac{2}{b}\right)\right\}  .\ \ \ \ \ ({\rm AdS_5})
\end{eqnarray}
We use the values of $\Delta$ from section \ref{subsec:mass:gf} to rewrite these expressions in terms of the weak mixing angle $s_Z$ rather than the bare angle $s_0$:
\begin{eqnarray}
 g_{ZWW} &=& e\frac{c_Z}{s_Z}
  \left[1 -
  \lambda\ \frac{(7+\kappa) - 6 c^2(1-\kappa)}{12 c^2 (c^2-s^2)
  (1+\kappa)} \right]~, \ \ \ \ \ {\rm (flat,\ brane\ localized)}\\
   g_{ZWW} &=& e\frac{c_Z}{s_Z}
  \left[1 + 
  \frac{\lambda}{12 c^2}\ \frac{7+\kappa}{1+\kappa} \right]~, \ \ \ {\rm (flat,\ ideal)}\\
g_{ZWW} &=& e\frac{c_Z}{s_Z}
  \left[1+\frac{1}{24}\frac{1}{c^2(c^2-s^2)}
   \frac{(10c^2-14)-\kappa(10 c^2+4)}{(1+\kappa)^2}
  \left(\frac{2}{b}\right)\right]~,\\
  && \hspace{8cm} {\rm (AdS_5,\ brane\ localized)}\nonumber\\
   g_{ZWW} &=& e\frac{c_Z}{s_Z}
  \left[1 + 
  \frac{1}{12 c^2}\ \frac{7+2\kappa}{(1+\kappa)^2}\left(\frac{2}{b}\right) \right]~. \ \ \ {\rm (AdS_5,\ ideal)}
\end{eqnarray}
The more compact notation $c$ or $s$ is used in an expression where the difference between employing the bare and corrected weak angles would cause only higher-order corrections.

Comparing these results with the form of eqn. (\ref{eq:tgvlag}) in section \ref{subsec:vertices:lang}, we see immediately that 
\begin{eqnarray}
\Delta g_1^Z =\Delta\kappa_Z&=&  - \frac{\lambda}{12 c^2}\frac{1}{(c^2-s^2)} \frac{(7+\kappa) - 6 c^2(1-\kappa)}{ 
  (1+\kappa)}~, \ \ \ {\rm (flat,\ brane\ localized)}\\
\Delta g_1^Z =\Delta\kappa_Z&=&  \frac{\lambda}{12 c^2}\ \frac{7+\kappa}{1+\kappa}~, \ \ \ {\rm (flat,\ ideal)}\label{eq:g1z:f}\\
\Delta g_1^Z =\Delta\kappa_Z&=& \frac{1}{24}\frac{1}{c^2(c^2-s^2)}
   \frac{(10c^2-14)-\kappa(10 c^2+4)}{(1+\kappa)^2}
  \left(\frac{2}{b}\right)~,  \\
&&  \hspace{8cm} {\rm (AdS_5, brane\ localized)}\nonumber \\
\Delta g_1^Z =\Delta\kappa_Z&=&   \frac{1}{12 c^2}
   \frac{7+2\kappa}{(1+\kappa)^2}
  \left(\frac{2}{b}\right)~. \label{eq:g1z:w}\ \ \ {\rm (AdS_5,\ ideal)}
\end{eqnarray}
We will relate these results to the values of chiral Lagrangian parameters in section \ref{sec:chiral} and will comment on the implications of experimental limits on $\Delta g_1^Z$ and $\Delta\kappa_Z$ in section \ref{sec:colphen}.

\subsection{Quartic gauge boson vertices}
\label{subsec:vertices:22:warp}

The quartic $W$-boson coupling $g_{WWWW}$ is evaluated in flat space by the overlap integral 
\begin{equation}
 g_{WWWW} = \int_{0}^{\pi R}dz 
  \left\{ \frac{1}{g_{5A}^2} \left|\chi_{W}^A(z)\right|^4 + 
          \frac{1}{g_{5B}^2} \left|\chi_{W}^B(z)\right|^4 \ 
 \right\} + 
 \frac{1}{g_{0}^2} \left|\chi_{W}^A(0)\right|^4 
\end{equation}
and in warped space using 
\begin{equation}
 g_{WWWW} = \int_{R}^{R'}dz \left(\frac{R}{z}\right)  
  \left\{ \frac{1}{g_{5A}^2} \left|\chi_{W}^A(z)\right|^4 + 
          \frac{1}{g_{5B}^2} \left|\chi_{W}^B(z)\right|^4 \ 
 \right\} .
\end{equation}
By inserting the appropriate forms of the  $W$ boson wavefunctions from section \ref{subsec:mass:wz}, we find
\begin{eqnarray}
 g_{WWWW} &=&\frac{e^2}{s_0^2}\left[ 
1+ \lambda \left\{ s^2\frac{(1+\kappa)^2}{\kappa}
	    -
	    \frac{7+38\kappa+52\kappa^2+15\kappa^3}{15(1+\kappa)^2}\right\}  
\right] \ \ \ \ {\rm (flat)}\\
g_{WWWW} &=& \frac{e^2}{s_0^2}\left[ 1+
			      \frac{1}{24}\frac{(-9\kappa^3+20\kappa+11)}{(1+\kappa)^4}\left(\frac{2}{b}\right)
			     \right] .\ \ \ \ ({\rm AdS_5})
\end{eqnarray}
We use the values of $\Delta$ from section \ref{subsec:mass:gf} to rewrite these expressions in terms of the weak mixing angle $s_Z$ rather than the bare angle $s_0$:
\begin{eqnarray}
 g_{WWWW} &=& \frac{e^2}{s_Z^2}\left[ 
1+ \lambda\ \frac{(-18+16c^2) + (-27+14c^2)\kappa - (3+14c^2)\kappa^2}{15
(c^2-s^2)\ (1+\kappa)^2}
\right] \\ & &\hspace{8cm}{\rm (flat,\ brane\ localized)}\nonumber,\\
  g_{WWWW} &=&\frac{e^2}{s_Z^2}\left[ 
1+ \frac{\lambda}{5}\ \frac{6 + 9\kappa +\kappa^2}{(1+\kappa)^2}
\right] \ \ \ \ \ \ \ \ \ \ \ \ \ \ \ \ {\rm (flat,\ ideal)}\\
   g_{WWWW} &=&\frac{e^2}{s_Z^2}\left[ 1 -
			      \frac{3}{4}\frac{1}{(1+\kappa)}\frac{1}{c^2-s^2}\left(\frac{2}{b}\right)
			      + 
			      \frac{1}{24}\frac{(-9\kappa^3+20\kappa+11)}{(1+\kappa)^4}\left(\frac{2}{b}\right)  
			     \right] \\ & &\hspace{8cm}{\rm (AdS_5,\ brane\ localized)}\nonumber,\\
    g_{WWWW} &=&\frac{e^2}{s_Z^2}\left[ 1 +
			      \frac{3}{4}\frac{1}{(1+\kappa)}\left(\frac{2}{b}\right)
			      + 
			      \frac{1}{24}\frac{(-9\kappa^3+20\kappa+11)}{(1+\kappa)^4}\left(\frac{2}{b}\right)  
			     \right] ~,\\
			     && \hspace{8cm} {\rm (AdS_5,\ ideal)}~.\nonumber
\end{eqnarray}
We will relate these results to the values of chiral Lagrangian parameters in section \ref{sec:chiral} and will comment on the phenomenological implications for $W$-boson scattering in section \ref{sec:concl}.


\section{Chiral Lagrangian Parameters}
\label{sec:chiral}

In studying the phenomenology of our models, it is useful to 
make contact with the parameterization afforded by the effective electroweak chiral Lagrangian.  Of the complete set of 12 CP-conserving operators written down by Longhitano \cite{Appelquist:1980vg,Longhitano:1980iz,Longhitano:1980tm,Appelquist:1980ix,Holdom:1990xq,Falk:1991cm} and Appelquist and Wu \cite{Appelquist:1993ka}, those which apply to our Higgsless models\footnote{In our models, weak isospin violation arises only through hypercharge and vanishes as $g_Y \to 0$.  Hence, the coefficients $\beta_1$ and $\alpha_{6...11}$ of the chiral Lagrangian operators that include weak isospin violation which persists in the limit $g_Y \to 0$ must vanish at tree-level.}  are the following:
\begin{eqnarray}
{\cal L}_1 &\equiv& \frac12 \alpha_1 g_W g_Y B_{\mu\nu} Tr(T W^{\mu\nu})\\
{\cal L}_2 &\equiv& \frac12 i \alpha_2 g_Y B_{\mu\nu} Tr(T [V^\mu, V^\nu])\\
{\cal L}_3 &\equiv& i \alpha_3 g_W Tr(W_{\mu\nu} [V^\mu, V^\nu])\\
{\cal L}_4 &\equiv& \alpha_4 [ Tr(V^\mu V^\nu)]^2\\
{\cal L}_5 &\equiv& \alpha_5 [Tr(V_\mu V^\mu)]^2~.
\end{eqnarray}
Here $W_{\mu\nu}$, $B_{\mu\nu}$, $T \equiv U\tau_3U^\dagger$ and $V_\mu \equiv (D_\mu U)U^\dagger$, with
$U$ being the nonlinear sigma-model field\footnote{$SU(2)_W \equiv SU(2)_L$ and $U(1)_Y$ is
identified with the $T_3$ part of $SU(2)_R$.}  arising from $SU(2)_L \otimes SU(2)_R \to
SU(2)_V$, 
are the $SU(2)_W$-covariant and $U(1)_Y$-invariant building blocks of the expansion.
An alternative parametrization by Gasser and Leutwyler \cite{Gasser:1983yg} gives a different set of names to the coefficients of these operators
\begin{equation}
\alpha_1 = L_{10}~,\ \ \ \ \ \ \ \ \ \ \alpha_2 = - \frac12 L_{9R}~,\ \ \ \ \ \ \ \ \ \  \alpha_3 = - \frac12 L_{9L}~,
\end{equation}
\begin{equation}
\alpha_4 = L_2~, \ \ \ \ \ \ \ \ \ \ \alpha_5 = L_1~.
\end{equation}

The chiral Lagrangian coefficients are related to $\alpha S$ and the Hagiwara-Peccei-Zeppenfeld-Hikasa \cite{Hagiwara:1986vm} triple-gauge-vertex parameters and the quartic $W$ boson vertex as follows:
\begin{eqnarray}
\alpha S &=& -16\pi\alpha (\alpha_1)~,\\
 \Delta g_1^Z &=& \frac{1}{c^2(c^2-s^2)}e^2\alpha_1 +
  \frac{1}{s^2c^2}e^2\alpha_3~,\\
 \Delta \kappa_Z &=&
  \frac{2}{c^2-s^2}e^2\alpha_1
  -\frac{1}{c^2}e^2\alpha_2+\frac{1}{s^2}e^2\alpha_3 , \\  
 \Delta \kappa_\gamma &=& \frac{1}{s^2}(-e^2\alpha_1 +e^2\alpha_2
  +e^2\alpha_3)~,\\
  g_{WWWW} &=& \frac{e^2}{s_Z^2} \left[1+\frac{2}{c^2-s^2}e^2\alpha_1 +
			       \frac{2}{s^2} e^2\alpha_3 + \frac{1}{s^2}
			       e^2\alpha_4\right] ~.
\end{eqnarray}
Again, in any vector-resonance model all multi-gauge vertices arise from re-expressing
the nonabelian couplings in the kinetic
energy terms in the original Lagrangian in terms of the mass-eignestate fields. 
In this case, we find that $\alpha_4 \equiv -\alpha_5$. 
The leading corrections to $WW$ and $WZ$ elastic scattering arise from $\alpha_{4,5}$
\cite{Chivukula:inprep}.


\begin{table}[ht]
\begin{center}
 \begin{tabular}{|c|c|c|}
 \hline
 flat $SU(2) \times SU(2)$ & & \\
 Longhitano parameters & brane localized & ideally delocalized \\
\hline
$e^2 \alpha_1$ & 
$- \frac{2}{3}\, \lambda\, s^2$ &
0\\
\hline
$e^2 \alpha_2$ & 
$  - \frac{1}{12} \left(\frac{7+\kappa}{1+\kappa}\right) \lambda\, s^2 $ &
$  - \frac{1}{12} \left(\frac{7+\kappa}{1+\kappa}\right) \lambda\, s^2 $ \\
\hline
$e^2 \alpha_3$ & 
$  - \frac{1}{12} \left(\frac{1+7\kappa}{1+\kappa}\right) \lambda\, s^2 $ &
$  \frac{1}{12} \left(\frac{7+\kappa}{1+\kappa}\right) \lambda\, s^2 $ \\
\hline
$e^2 \alpha_4$ & 
$ \frac{1}{30}\, \frac{(1+14\kappa+\kappa^2)}{(1+\kappa)^2}\, \lambda\, s^2\ $ &
$ \frac{1}{30}\, \frac{(1+14\kappa+\kappa^2)}{(1+\kappa)^2}\, \lambda\, s^2\ $
\\
\hline
$e^2 \alpha_5$ & 
$-  \frac{1}{30}\, \frac{(1+14\kappa+\kappa^2)}{(1+\kappa)^2}\, \lambda\, s^2\ $&
$ - \frac{1}{30}\, \frac{(1+14\kappa+\kappa^2)}{(1+\kappa)^2}\, \lambda\, s^2\ $
\\
  \hline
 \end{tabular}
\end{center}
\caption{Longhitano's parameters in $SU(2)\otimes SU(2)$
 flat Higgsless models for the cases of brane localized fermions and
 ideally delocalized fermions. }
 \label{table:SU2SU2:f}
\end{table}


\begin{table}[ht]
\begin{center}
 \begin{tabular}{|c|c|c|}
 \hline
 ${\rm AdS_5}$ $SU(2) \times SU(2)$ & & \\
 Longhitano parameters  
             &   brane localized  & ideally delocalized  \\
  \hline
   $e^2\alpha_1$  &   $-\frac{3s^2}{4}\frac{1}{(1+\kappa)}\left(\frac{2}{b}\right) $ & 0 \\
  \hline
   $e^2\alpha_2$  &  $ -\frac{s^2}{12}\frac{7+2\kappa}{(1+\kappa)^2}\left(\frac{2}{b}\right)$& $ -\frac{s^2}{12}\frac{7+2\kappa}{(1+\kappa)^2}\left(\frac{2}{b}\right)$\\
  \hline
   $e^2\alpha_3$  & $-\frac{s^2}{12}\frac{2+7\kappa}{(1+\kappa)^2}
 \left(\frac{2}{b}\right) $& 
$ \frac{s^2}{12}\frac{7+2\kappa}{(1+\kappa)^2}\left(\frac{2}{b}\right)$\\
  \hline
   $e^2\alpha_4$  & $
  \frac{s^2}{24}\frac{ 1 + 9\kappa + \kappa^2 }{(1+\kappa)^3} 
  \left(\frac{2}{b}\right) $& $
  \frac{s^2}{24}\frac{ 1 + 9\kappa + \kappa^2 }{(1+\kappa)^3} 
  \left(\frac{2}{b}\right) $\\
  \hline
   $e^2\alpha_5$  & $-
  \frac{s^2}{24}\frac{ 1 + 9\kappa + \kappa^2 }{(1+\kappa)^3} 
  \left(\frac{2}{b}\right) $ & $-
  \frac{s^2}{24}\frac{ 1 + 9\kappa + \kappa^2 }{(1+\kappa)^3} 
  \left(\frac{2}{b}\right) $\\
  \hline
 \end{tabular}
\end{center}
\caption{Longhitano's parameters in $SU(2)\otimes SU(2)$
 warped Higgsless models for the cases of the brane localized fermions
 and the ideally delocalized fermions.}
\label{table:SU2SU2:w}
\end{table}


\begin{table}[ht]
\begin{center}
 \begin{tabular}{|c|c|c|}
   \hline
 ${\rm AdS_5}$ $SU(2)$ & & \\
Longhitano parameters
             &   brane localized  & ideally delocalized  \\
  \hline
   $e^2\alpha_1$  & $-\frac{s^2}{4}\left(\frac{2}{b}\right)$ & $0$ \\
  \hline
   $e^2\alpha_2$  & $-\frac{s^2}{12}\left(\frac{2}{b}\right)$ &$-\frac{s^2}{12}\left(\frac{2}{b}\right)$  \\
  \hline
   $e^2\alpha_3$  & $-\frac{s^2}{6}\left(\frac{2}{b}\right)$ & $\frac{s^2}{12}\left(\frac{2}{b}\right)$ \\
  \hline
   $e^2\alpha_4$  & $ \frac{s^2}{24}\left(\frac{2}{b}\right)$ & 
   $ \frac{s^2}{24}\left(\frac{2}{b}\right)$ \\
  \hline
   $e^2\alpha_5$  & $-\frac{s^2}{24}\left(\frac{2}{b}\right)$ & 
   $-\frac{s^2}{24}\left(\frac{2}{b}\right)$ \\
  \hline
 \end{tabular}
\end{center}
\caption{Longhitano's parameters in $SU(2)$  warped Higgsless
 models for the cases of brane localized fermions and 
ideally delocalized fermions.}
\label{table:SU2:w}
\end{table}


Using these relationships and the values previously derived for $\alpha S$, $\Delta g_1^Z$, $\Delta\kappa_Z$, $\Delta \kappa_\gamma$, and $g_{WWWW}$, we arrive at values for the $\alpha_i$ in each of the $SU(2)_A \times SU(2)_B$ Higgsless models we have been considering.  The values are given in tables \ref{table:SU2SU2:f} and \ref{table:SU2SU2:w} .   These values are consistent with several symmetry considerations.  First, $\alpha_2 \equiv -  L_{9R}/2$ is the coefficient of an operator that is not related to the $SU(2)_W$ properties of the model; as such, this coefficient should be unaffected by the degree of delocalization of the $SU(2)_W$ properties of the fermions.  Indeed, we see that $\alpha_2$ is the same for both the brane-localized and ideal fermions.  Conversely, we expect the values of $\alpha_1$ and $\alpha_3$ to be sensitive to the $SU(2)_W$ properties of the fermions and this is observed in our results, yielding an examples theories in which $\alpha_2 \neq \alpha_3$
(or $L_{9R} \neq L_{9L}$).   Third, in the limit where $\kappa \to 1$, the models with brane-localized fermions should display an $A \leftrightarrow B$ parity; this is consistent with the fact that $\alpha_2 = \alpha_3$ for $\kappa = 1$.  Finally, since $\Delta\kappa_\gamma\equiv 0$,
we find  $\alpha_2 = - \alpha_3$ for the case of ideal delocalization, in which 
$\alpha_1=0$.

We have also evaluated the chiral Lagrangian parameters for $SU(2)$ Higgsless models.  
 A flat space $SU(2)$ model defined on the interval $0 \leq z \leq 2\pi R$ is simply the  $\kappa = 1$ limit of the $SU(2)_A \times SU(2)_B$ model considered here; the values of the $\alpha_i$ may be read fairly easily from Table \ref{table:SU2SU2:f}.   The $\alpha_i$ for a warped-space $SU(2)$ model are given in table \ref{table:SU2:w}.  The points where the derivation of results for the warped-space $SU(2)$ model differs from the analysis of $SU(2)_A \times SU(2)_B$ are covered in Appendix \ref{appendix2}.


\section{Collider Phenomenology}
\label{sec:colphen}

\subsection{Triple Gauge Vertices}
\label{subsec:colphen1}

For a model with brane-localized fermions, the non-zero value of $\alpha S$ provides
a strong constraint on the mass of the lightest $KK$ resonance \cite{SekharChivukula:2004mu}.  However, in models with ideally-delocalized fermions, $\alpha S = 0$.  In these models, experimental constraints from measurements of the triple-gauge-boson vertices can provide valuable bounds on the $KK$ masses.  

Currently, the strongest bounds on $\Delta g^Z_1$ and $\Delta \kappa_Z$ come from LEP II.
 The 95\% c.l. upper limit (recalling that $\Delta g^Z_1$ is positive in our models) is $\Delta g^Z_1 \leq 0.028$ \cite{LEPEWWG}.
We can estimate the degree to which this constrains Higgsless models with ideal delocalizaation by considering how $\Delta g^Z_1$ is related to the mass of the lightest $KK$ resonance.  

For an $SU(2)_A \times SU(2)_B$ flat-space model, the form of $\Delta g_1^Z$ is shown in  
eqn. (\ref{eq:g1z:f}) to depend on $\lambda$ .  From equations (\ref{eq:fmwi})
and (\ref{eq:fmwii}), we see that $\lambda \approx (\pi R M_W)^2$; furthermore, as just discussed above, the mass of the lightest $KK$ resonance is $M_{W1} \approx 1/2R$, independent\footnote{In the limit where $g_0 = 0$ it is straightforward to see that there are two ways for a $KK$ resonance profile to satisfy the boundary conditions at $z=\pi R$ (flat) or $z = R'$ (warped).  Either the profile can vanish at the boundary, or its z-derivitive can vanish there.  In neither case does the mass of the resonance depend on $\kappa$.} of $\kappa$, so we find that
\begin{equation}
\Delta g^Z_1 = \frac{\pi^2}{12 c^2} \left(\frac{M_{W}}{M_{W_1}}\right)^2 \left[\frac14 \cdot \frac{7 + \kappa}{1+\kappa}\right]~
\end{equation}
where the factor in square brackets equals 1 for $\kappa=1$.  In an $SU(2)_A \times SU(2)_B$ model in $AdS_5$ space, eqn. (\ref{eq:g1z:w}) shows that $\Delta g^Z_1$ depends on  $b$.  In this model, $1/b \approx \frac14 (1 + \kappa) (M_{W} R')^2$ and $R' = x_1 / M_{W_1}$ (again, independent$^{**}$ of $\kappa$) where $x_1\approx 2.4$ is the first zero of the Bessel function $J_0$.   Putting this all together, we have
\begin{equation}
\Delta g^Z_1 = \frac{3 x_1^2}{16 c^2}  \left(\frac{M_{W}}{M_{W_1}}\right)^2 \left[\frac29 \cdot  \frac{7 + 2\kappa}{1 + \kappa}\right]
\end{equation}
where the factor in square brackets equals 1 for $\kappa=1$.

Inserting numerical values for $M_W$, $c$, and $y_1$ and denoting the 95\% c.l. experimental 
upper bound on $\Delta g^Z_1$ as $\Delta g_{max}$, we find the bound
\begin{eqnarray}
M_{W_1} &\geq& \sqrt{\frac{6900}{\Delta g_{max}} \ \left[\frac14 \cdot \frac{7 + \kappa}{1+\kappa}\right]}\ {\rm GeV}~, \ \ \ \ \ {\rm (flat)}\\
M_{W_1} &\geq& \sqrt{\frac{9100}{\Delta g_{max}} \ \left[\frac29 \cdot  \frac{7 + 2\kappa}{1 + \kappa}\right]} \  {\rm GeV}~. \ \ \ \ \ (AdS_5)
\end{eqnarray}
The LEP II data therefore implies a 95\% c.l. lower bound of 500 (570) GeV on the first $KK$ resonance in flat (warped) space models for $\kappa = 1$ and lower bounds of 250 - 650 (380 - 700) GeV as $\kappa$ varies from $\infty$ to 0.

Future experiments will improve the limits on $\Delta g_1^Z$ and $\Delta \kappa_Z$ significantly.
An analysis of $WZ$ production at the LHC by Dobbs \cite{Dobbs:2005ev} including both systematic and statistical effects finds that with 30 $fb^{-1}$ of integrated luminosity it should possible to set a 95\% c.l. bound of $-0.0086 < \Delta g^Z_1 < 0.011$. The limit on $\Delta\kappa_Z$ is expected to be significantly looser, as are likely limits from single electroweak gauge boson production at LHC \cite{Eboli:2004gc}.  It therefore appears that LHC would be sensitive to $KK$ resonances up to 790 GeV for a flat-space model and 960 GeV for an $AdS_5$  space model for $\kappa = 1$.  Reference \cite{Menges:2001gg} finds that a linear electron-positron collider with polarized beams should be sensitive to both $\Delta g^Z_1$ and $\Delta\kappa_Z$.  The anticipated $2\sigma$ upper bounds on $\Delta g^Z_1$ are 0.0048 (0.0027) for a 500 GeV (800 GeV) collider; for $\Delta\kappa_Z$ the anticipated limit is 0.00098 (0.00042) for a 500 GeV (800 GeV) collider.   Thus, a 500 GeV (800 GeV) linear collider would be sensitive to a $KK$ resonance of mass up to 2.6 TeV (4 TeV) in a flat space model and up to 3.2 TeV (4.9 TeV) in a warped space model for $\kappa = 1$. 

A similar analysis reveals that neither the LHC nor a linear collider \cite{Boos:1999kj} 
will be able to probe the quartic gauge boson vertices per se because the masses of the KK resonances 
required to yield visible values of $\alpha_4=-\alpha_5$ are similar in magnitude 
to the relevant sub-process scattering energies.  Rather, one
would look directly for resonances in elastic $WW$ or $WZ$ scattering (as discussed briefly below).

\subsection{Direct Searches}
\label{subsec:colphen2}

It is interesting to note Higgsless models with ideally-delocalized fermions cannot be constrained by  direct collider searches for new gauge bosons that rely on the bosons' couplings to fermions.  As we have seen, the higher $KK$ resonances are fermiophobic to leading order (their coupling to fermions is suppressed by $M_W^2 / M_{W_{(n)}}^2$).  
 
Existing searches for $W'$ bosons \cite{PDG} assume that the $W'$ bosons couple to ordinary quarks and leptons (generally with SM strength). All assume that the $W'$ is produced via these couplings; all but one also assume that the $W'$ decays only to fermions and that the $W' \to W Z$ decay channel is  unavailable.   However, by construction, the ideally delocalized fermions in our Higgsless models have no coupling to the $KK$ resonances $W_{(n\geq1)}$.  Hence, none of the current searches apply.   Proposed future searches via the Drell-Yan process at the LHC,  or via $e^+ e^- \to \nu \bar\nu \gamma$ \cite{Godfrey:2000hc} and $e \gamma \to \nu q + X$ \cite{Godfrey:2000pw} at a linear collider also rely on $W'$ couplings to fermions and will not apply either.  

Likewise, existing direct searches for $Z'$ bosons \cite{PDG} rely on $Z' f \bar f$ couplings and assume that the decay channel $Z' \to W^+W^-$ is not available; as such, they do not constrain the  $Z_{(n\geq 1)}$ of our Higgsless models with ideally delocalized fermions.  Proposed future searches at LHC and the linear collider that rely on fermionic couplings for production or decay of the $Z'$ will not apply either.  
 
The only way to perform a direct search for the $W_{(n\geq1)}$ and $Z_{(n\geq 1)}$ states will be to study $WW$ or $WZ$ elastic scattering.   If no resonances are seen, these processes will also 
afford the opportunity to constrain the values of the chiral Lagrangian parameters $\alpha_4$ and $\alpha_5$.  We will comment further on vector boson scattering in Higgsless models in future work \cite{Chivukula:inprep}.


\section{Conclusions}
\label{sec:concl}

In this paper we have extended
the analysis of Higgsless models with ideal delocalization in several ways.   
We have computed the form of the triple and quartic gauge
boson vertices in these models and related them to the parameters of the electroweak chiral Lagrangian.  As contributions to the $S$ parameter vanish to leading order, current constraints
on these models arise from limits on deviations in multi-gauge-boson vertices and these constraints were shown to provide lower 
bounds of order a few hundred GeV on the masses of the lightest $KK$ resonances above the 
$W$ and $Z$ bosons.  We also studied the collider phenomenology of the $KK$ resonances in 
models with ideal delocalization. We showed that these resonances are 
fermiophobic -- therefore traditional direct collider searches are not sensitive to them and measurements of gauge-boson scattering will be needed to find them.

\acknowledgments

R.S.C. and E.H.S. are supported in part by the US National Science Foundation under
grant  PHY-0354226. M.K. is supported by a MEXT Grant-in-Aid for Scientific Research
No. 14046201.
M.T.'s work is supported in part by the JSPS Grant-in-Aid for Scientific Research No.16540226. H.J.H. is supported by the US Department of Energy grant
DE-FG03-93ER40757.  R.S.C., M.K., E.H.S., and M.T. gratefully acknowledge the hospitality of the Aspen Center for Physics where this work was completed.


\appendix
\section{Normalization of fermion wavefunction}
\label{appendix1}

The 5D action of the $\partial^\mu \gamma_\mu$ portion of the
fermionic kinetic term takes the following form \cite{Csaki:2003sh}:
\begin{equation}
 i\,\int d^4x\, dz N^{\psi}(z)\, {\bar \Psi(x,z)} \gamma^\mu \partial_\mu \Psi(x,z),
\end{equation}
where $N^{\psi}(z)$ is the factor which is determined from $\sqrt{g}$ 
and the ``f\"{u}nfbein" in the specific metric. 
($N^{\psi}(z)=\left(\frac{R}{z}\right)^4$ for the case of warped metric.)

Gauge invariance requires that the coupling of fermionic current 
to the gauge boson $W_\mu(x,z)$ takes the form of
\begin{equation}
 \int d^4x dz N^{\psi}(z) {\bar \Psi(x,z)} \gamma^\mu W_\mu(x,z) \Psi(x,z) .
\end{equation}
The $KK$-decomposition of $W_\mu(x,z)$ and $\Psi(x,z)$ are expressed as follows:
\begin{eqnarray}
 W_\mu(x,z) &=& \sum_n W_\mu^{(n)}(x)\,
  \chi_{W_{(n)}}(z) ,   \\  
 \Psi(x,z) &=& \sum_n \psi^{(n)}(x)\,
  \chi_{\psi_{(n)}}(z) . 
\end{eqnarray}
If we require the 4D fermion kinetic energy to be canonical, 
$\chi_{\psi_{(n)}}(z)$ must be normalized by the following condition:
\begin{equation}
 \int dz N^{\psi}(z) \left|\chi_{\psi_{(n)}}(z)\right|^2 = 1.
\label{eq:norm}
\end{equation}
In this case, couplings of the fermion zero-mode $\psi^{(0)}(x)$ 
to $W_\mu^{(n)}$ bosons are expressed as follows:
\begin{equation}
 g_{W_{(n)}} = \int dz N^{\psi}(z) \left|\chi_{\psi_{(0)}}(z)\right|^2 
\chi_{W_{(n)}}.
\label{eq:coupling}
\end{equation}
Since weight function $N^{\psi}(z)$ appears both in Eq.~(\ref{eq:norm}) 
and Eq.~(\ref{eq:coupling}), it is convenient to include $N^{\psi}(z)$ 
in the definition of fermion wavefunction:
\begin{equation}
 \left|\psi(z)\right|^2 \equiv N^{\psi}(z) \left|\chi_{\psi_{(0)}}(z)\right|^2.
\end{equation}
Then, the normalization condition and couplings are expressed 
as follows:
\begin{equation}
 \int dz \left|\psi(z)\right|^2 = 1,
\label{eq:norm2}
\end{equation}
\begin{equation}
 g_{W_{(n)}} = \int dz  \left|\psi(z)\right|^2 
\chi_{W_{(n)}}.
\label{eq:coupling2}
\end{equation}

The orthonormal conditions for $\chi_{W_{(n)}}$ are given by
\begin{equation}
 \int dz N^{W}(z) \chi_{W_{(n)}} \chi_{W_{(m)}} = \delta_{n,m},
\end{equation}
therefore the  profile of ideally delocalized fermion is expressed as :
\begin{equation}
 \left|\psi_{\rm ideal}(z)\right|^2 \propto N^{W}(z) \chi_{W_{(0)}}(z),
\end{equation}
where the overall normalization constant is determined from
Eq.~(\ref{eq:norm2}) . 



\section{$SU(2)$ Model in $AdS_5$}
\label{appendix2}

This appendix sketches the analysis of the $SU(2)$ Higgsless modes in warped
space, focusing on where it differs from that of the $SU(2)\times SU(2)$ models discussed in the text.  
We start from the 5D action of an $SU(2)$ gauge theory, using the $AdS_5$ metric of eq. (\ref{eq:adsmetric})
\begin{equation}
  S_{5D} = \int_{R}^{R'} dz \dfrac{R}{z\, g_5^2} \int d^4 x \left[
    -\frac{1}{4} \eta^{\mu\alpha}\eta^{\nu\beta} 
        W^a_{\mu\nu} W^a_{\alpha\beta}
    +\frac{1}{2} \eta^{\mu\nu} W^a_{\mu z} W^a_{\nu z} 
  \right].
\end{equation}
The boundary conditions are taken as
\begin{equation}
  \left. \partial_z W^{1,2,3}_{\mu}(x,z) \right|_{z=R}=0, \qquad
  \left. W^{1,2}_{\mu}(x,z) \right|_{z=R'}=0, \qquad
  \left. \partial_z W^{3}_{\mu}(x,z) \right|_{z=R'}=0~,
\end{equation}
in order to make the theory consistent with the standard model
symmetry breaking structure
$  SU(2)\times U(1) \to U(1)$.
To ensure a non-trivial weak mixing angle, we further introduce
a brane kinetic term at $z=R'$,
\begin{equation}
  S_{\rm TeV} = \int_R^{R'} dz \dfrac{1}{g_Y^2} \delta(z-R'+\epsilon)
  \int d^4 x \left[
    -\frac{1}{4} \eta^{\mu\alpha} \eta^{\nu\beta} 
     W^3_{\mu\nu} W^3_{\alpha\beta}
  \right], \qquad 
  (\epsilon \to 0+).
\label{eq:brane_kinetic}
\end{equation}
We also assume that the  $U(1)_Y$ fermion coupling is localized at
the $z=R'$ brane.

The 5D field $W_\mu(x,z)$ can be decomposed into $KK$-modes, whose
mode functions $\chi$ obey differential equations and $z=R$ boundary conditions
identical to those of $\chi_W^A$ in the $SU(2)\times SU(2)$ model.  The $z=R'$ boundary conditions 
\begin{equation}
  0 = \left. \chi_{W(n)}(z) \right|_{z=R'}, \quad
  0 = \left. \partial_z \chi_\gamma(z) \right|_{z=R'}, \quad
  0 = \left. \partial_z \chi_{Z(n)}(z) 
        -\dfrac{R'}{R} \dfrac{g_5^2}{g_Y^2} M_{Z(n)}^2
         \chi_{Z(n)}(z) 
      \right|_{z=R'}~,
\label{eq:BC_Rp}
\end{equation}
reflect the hypercharge brane kinetic term.  

The expressions for $e$, $\chi_\gamma$ and $s_0$ are obtained just as in the main text, and differ
only in the absence of terms involving $g_{5B}^2$ .  

The $W$ boson mass is found to be 
\begin{equation}
  (M_W R')^2 = \dfrac{4}{b}\left(1 + \dfrac{3}{2b}\right)
               + {\cal O}\left(\frac{1}{b^3}\right).
\end{equation}
and the profile is 
\begin{equation}
  \chi_W(z) = \frac{e}{s_0} \left[1+\frac{3}{8}\frac{2}{b}\right]
  \left[
  1 - \frac{1}{2} \left(\ln\dfrac{z}{R}-\frac{1}{2}\right)(M_W z)^2
    +\frac{1}{16} \left(\ln\dfrac{z}{R}\right)(M_W z)^4
  \right],
\label{eq:W_profile}
\end{equation}
Note that, as one would expect, this mass and profile agree with those
of $W^A$ in the $SU(2)\times SU(2)$ model when one sets $\kappa = 0$.
The mass and profile of the  $Z$ boson can likewise be obtained:
\begin{equation}
  (M_Z R')^2 = \frac{4}{b} \dfrac{1}{c_0^2}
          \left[
            1 + \frac{1}{2b} \left(3-\dfrac{s_0^2}{c_0^2} \right)
          \right]
         + {\cal O}\left(\frac{1}{b^3}\right),
\label{eq:M_Z}
\end{equation}
\begin{equation}
  \chi_Z(z) = e \frac{c_0}{s_0}
    \left[1+\frac{3}{8c_0^2}\frac{2}{b}\right]
  \left[
  1 - \frac{1}{2} \left(\ln\dfrac{z}{R}-\frac{1}{2}\right)(M_Z z)^2
    +\frac{1}{16} \left(\ln\dfrac{z}{R}\right)(M_Z z)^4
  \right].
\end{equation}
These reflect the influence a hypercharge brane
kinetic term at $z=R$ rather than at $z=R'$.

The calculations of the fermion profiles, $g_W$ and $G_F$ follow the route laid
out in the main text.   For brane-localized fermions, these quantities depend only on $\chi^A_W$ in the $SU(2)\times SU(2)$ model and taking the $\kappa=0$ limit yields their values in the $SU(2)$ model. For ideally-delocalized fermions, we find 
\begin{equation}
C_{ideal} = g_W = \left[4 \sqrt{2} G_F M_W^2\right]^{\frac12}= \frac{e}{s_0} \left(1 + \frac18 \frac2b\right)~.
\end{equation}
 The value of $\Delta$ relating $s_0$ to $s_Z$ is 
\begin{eqnarray}
  \Delta &=& -\frac{s^2}{4} \dfrac{3c^2-s^2}{c^2-s^2}\frac{2}{b}~, \ \ \ \ \ {\rm (brane\ localized)}\\
  \Delta &=& -\frac{s^2}{4} \frac{2}{b}~. \ \ \ \ \ {\rm (ideal)}
  \label{eq:thiss}
\end{eqnarray}

Performing the appropriate integrals yields the multi-gauge-boson vertex couplings
\begin{eqnarray}
g_{\gamma WW}^2 
  &=& e^2~, \\
  g_{ZWW}^2 
  &=& \frac{c_0^2}{s_0^2} e^2 \left[
         1 + \dfrac{5}{12c^2}\dfrac{2}{b} 
      \right]~,\\
       g_{WWWW} &=&  \dfrac{e^2}{s_0^2} \left[ 1 +\frac{11}{24}\frac{2}{b} \right]~.
\end{eqnarray}
and, after applying eqns. (\ref{eq:thiss}), the Hagiwara-Peccei-Zeppenfeld-Hikasa parameters emerge 
\begin{eqnarray}
  \Delta \kappa_\gamma &=& 0, \qquad
  \Delta g_1^Z = \Delta \kappa_Z = 
  - \dfrac{1+c^2}{12c^2(c^2-s^2)}\dfrac{2}{b},\ \ \ \ \ {\rm (brane\ localized)}\\
   \Delta \kappa_\gamma &=& 0, \qquad
  \Delta g_1^Z = \Delta \kappa_Z = 
  \dfrac{1}{12c^2}\dfrac{2}{b}. \ \ \ \ \ {\rm (ideal)}
\end{eqnarray}
From these, the values of the $\alpha_i$ listed in Table 3 are readily derived.  Note
that $\alpha_4 = -\alpha_5$ matches the $\kappa=0$ value for the warped $SU(2)\times SU(2)$ model because $g_{WWWW}$ depends only on $\chi_W^A$ in the $SU(2)\times SU(2)$ model at $\kappa=0$.



\begin{thebibliography}{99}

\bibitem{Csaki:2003dt}
  C.~Csaki, C.~Grojean, H.~Murayama, L.~Pilo and J.~Terning,
{\it Gauge theories on an interval: Unitarity without a Higgs},
  Phys.\ Rev.\ D {\bf 69}, 055006 (2004)
  [arXiv:hep-ph/0305237].

\bibitem{Higgs:1964ia}
P.~W. Higgs, {\it Broken symmetries, massless particles and gauge fields},
  {\em Phys. Lett.} {\bf 12} (1964) 132--133.


%
\bibitem{Agashe:2003zs}
  K.~Agashe, A.~Delgado, M.~J.~May and R.~Sundrum,
  {\it RS1, Custodial Isospin and Precision Tests},
  JHEP {\bf 0308}, 050 (2003)
  [arXiv:hep-ph/0308036].

\bibitem{Csaki:2003zu}
C.~Csaki, C.~Grojean, L.~Pilo, and J.~Terning, {\it Towards a realistic model
  of higgsless electroweak symmetry breaking},  {\em Phys. Rev. Lett.} {\bf 92}
  (2004) 101802, [\href{http://xxx.lanl.gov/abs/hep-ph/0308038}{{\tt
  hep-ph/0308038}}].
 
 \bibitem{SekharChivukula:2001hz}
R.~Sekhar~Chivukula, D.~A. Dicus, and H.-J. He, {\it Unitarity of compactified
  five dimensional yang-mills theory},  {\em Phys. Lett.} {\bf B525} (2002)
  175--182, [\href{http://xxx.lanl.gov/abs/hep-ph/0111016}{{\tt
  hep-ph/0111016}}].

\bibitem{Chivukula:2002ej}
R.~S. Chivukula and H.-J. He, {\it Unitarity of deconstructed five-dimensional
  yang-mills theory},  {\em Phys. Lett.} {\bf B532} (2002) 121--128,
  [\href{http://xxx.lanl.gov/abs/hep-ph/0201164}{{\tt hep-ph/0201164}}].

\bibitem{Chivukula:2003kq}
R.~S. Chivukula, D.~A. Dicus, H.-J. He, and S.~Nandi, {\it Unitarity of the
  higher dimensional standard model},  {\em Phys. Lett.} {\bf B562} (2003)
  109--117, [\href{http://xxx.lanl.gov/abs/hep-ph/0302263}{{\tt
  hep-ph/0302263}}].

\bibitem{He:2004zr}
H.-J.~He,
{\it Higgsless deconstruction without boundary condition},
arXiv:hep-ph/0412113.



\bibitem{Cacciapaglia:2004jz}
  G.~Cacciapaglia, C.~Csaki, C.~Grojean and J.~Terning,
  {\it Oblique corrections from Higgsless models in warped space},
  Phys.\ Rev.\ D {\bf 70}, 075014 (2004)
  [arXiv:hep-ph/0401160].

\bibitem{Nomura:2003du}
Y.~Nomura, {\it Higgsless theory of electroweak symmetry breaking from warped
  space},  {\em JHEP} {\bf 11} (2003) 050,
  [\href{http://xxx.lanl.gov/abs/hep-ph/0309189}{{\tt hep-ph/0309189}}].

\bibitem{Barbieri:2003pr}
R.~Barbieri, A.~Pomarol and R.~Rattazzi,
{\it Weakly coupled Higgsless theories and precision electroweak tests},
Phys.\ Lett.\ B {\bf 591} (2004) 141
[arXiv:hep-ph/0310285].

\bibitem{Davoudiasl:2003me}
H.~Davoudiasl, J.~L.~Hewett, B.~Lillie and T.~G.~Rizzo,
{\it Higgsless electroweak symmetry breaking in warped backgrounds:  constraints
and signatures},
Phys.\ Rev.\ D {\bf 70} (2004) 015006
[arXiv:hep-ph/0312193].

\bibitem{Burdman:2003ya}
G.~Burdman and Y.~Nomura,
{\it Holographic theories of electroweak symmetry breaking without a Higgs
boson},
Phys.\ Rev.\ D {\bf 69} (2004) 115013
[arXiv:hep-ph/0312247].

\bibitem{Davoudiasl:2004pw}
H.~Davoudiasl, J.~L. Hewett, B.~Lillie, and T.~G. Rizzo, {\it Warped higgsless
  models with ir-brane kinetic terms},  {\em JHEP} {\bf 05} (2004) 015,
  [\href{http://xxx.lanl.gov/abs/hep-ph/0403300}{{\tt hep-ph/0403300}}].

\bibitem{Barbieri:2004qk}
R.~Barbieri, A.~Pomarol, R.~Rattazzi and A.~Strumia,
{\it Electroweak symmetry breaking after LEP1 and LEP2},
Nucl.\ Phys.\ B {\bf 703} (2004) 127
[arXiv:hep-ph/0405040].

\bibitem{Hewett:2004dv}
J.~L. Hewett, B.~Lillie, and T.~G. Rizzo, {\it Monte carlo exploration of
  warped higgsless models},  {\em JHEP} {\bf 10} (2004) 014,
  [\href{http://xxx.lanl.gov/abs/hep-ph/0407059}{{\tt hep-ph/0407059}}].

 \bibitem{Foadi:2003xa}
R.~Foadi, S.~Gopalakrishna, and C.~Schmidt, {\it Higgsless electroweak symmetry
  breaking from theory space},  {\em JHEP} {\bf 03} (2004) 042,
  [\href{http://xxx.lanl.gov/abs/hep-ph/0312324}{{\tt hep-ph/0312324}}].

\bibitem{Hirn:2004ze}
  J.~Hirn and J.~Stern,
  {\it The role of spurions in Higgs-less electroweak effective theories},
  Eur.\ Phys.\ J.\ C {\bf 34}, 447 (2004)
  [arXiv:hep-ph/0401032].

\bibitem{Casalbuoni:2004id}
R.~Casalbuoni, S.~De Curtis and D.~Dominici,
{\it Moose models with vanishing S parameter},
Phys.\ Rev.\ D {\bf 70} (2004) 055010
[arXiv:hep-ph/0405188].

\bibitem{Chivukula:2004pk}
R.~S.~Chivukula, E.~H.~Simmons, H.~J.~He, M.~Kurachi and M.~Tanabashi,
{\it The structure of corrections to electroweak interactions in Higgsless
models},
Phys.\ Rev.\ D {\bf 70} (2004) 075008
[arXiv:hep-ph/0406077].


\bibitem{Perelstein:2004sc}
M.~Perelstein, {\it Gauge-assisted technicolor?},  {\em JHEP} {\bf 10} (2004)
  010, [\href{http://xxx.lanl.gov/abs/hep-ph/0408072}{{\tt hep-ph/0408072}}].

\bibitem{Chivukula:2004af}
R.~S. Chivukula, E.~H. Simmons, H.-J. He, M.~Kurachi, and M.~Tanabashi, {\it
  Universal non-oblique corrections in higgsless models and beyond},  {\em
  Phys. Lett.} {\bf B603} (2004) 210--218,
  [\href{http://xxx.lanl.gov/abs/hep-ph/0408262}{{\tt hep-ph/0408262}}].

\bibitem{Georgi:2004iy}
  H.~Georgi, {\it Fun with Higgsless theories},
  Phys.\ Rev.\ D {\bf 71}, 015016 (2005)
  [arXiv:hep-ph/0408067].

\bibitem{SekharChivukula:2004mu}
R.~Sekhar Chivukula, E.~H.~Simmons, H.~J.~He, M.~Kurachi and M.~Tanabashi,
{\it Electroweak corrections and unitarity in linear moose models},
Phys.\ Rev.\ D {\bf 71} (2005) 035007
[arXiv:hep-ph/0410154].

\bibitem{Arkani-Hamed:2001ca}
N.~Arkani-Hamed, A.~G. Cohen, and H.~Georgi, {\it (de)constructing dimensions},
   {\em Phys. Rev. Lett.} {\bf 86} (2001) 4757--4761,
  [\href{http://xxx.lanl.gov/abs/hep-th/0104005}{{\tt hep-th/0104005}}].

\bibitem{Hill:2000mu}
C.~T. Hill, S.~Pokorski, and J.~Wang, {\it Gauge invariant effective lagrangian
  for kaluza-klein modes},  {\em Phys. Rev.} {\bf D64} (2001) 105005,
  [\href{http://xxx.lanl.gov/abs/hep-th/0104035}{{\tt hep-th/0104035}}].

\bibitem{Peskin:1992sw}
M.~E. Peskin and T.~Takeuchi, {\it Estimation of oblique electroweak
  corrections},  {\em Phys. Rev.} {\bf D46} (1992) 381--409.

\bibitem{Altarelli:1990zd}
G.~Altarelli and R.~Barbieri, {\it Vacuum polarization effects of new physics
  on electroweak processes},  {\em Phys. Lett.} {\bf B253} (1991) 161--167.

\bibitem{Altarelli:1991fk}
G.~Altarelli, R.~Barbieri, and S.~Jadach, {\it Toward a model independent
  analysis of electroweak data},  {\em Nucl. Phys.} {\bf B369} (1992) 3--32.

\bibitem{Georgi:1985hf}
H.~Georgi, {\it A tool kit for builders of composite models},  {\em Nucl.
  Phys.} {\bf B266} (1986) 274.

\bibitem{Cacciapaglia:2004rb}
G.~Cacciapaglia, C.~Csaki, C.~Grojean and J.~Terning,
{\it Curing the ills of Higgsless models: The S parameter and unitarity},
Phys.\ Rev.\ D {\bf 71} (2005) 035015
[arXiv:hep-ph/0409126].

\bibitem{Foadi:2004ps}
R.~Foadi, S.~Gopalakrishna and C.~Schmidt,
{\it Effects of fermion localization in Higgsless theories and electroweak
constraints},
Phys.\ Lett.\ B {\bf 606} (2005) 157
[arXiv:hep-ph/0409266].

\bibitem{Cacciapaglia:2005pa}
  G.~Cacciapaglia, C.~Csaki, C.~Grojean, M.~Reece and J.~Terning,
 {\it Top and bottom: A brane of their own},
  arXiv:hep-ph/0505001.

\bibitem{Chivukula:2005bn}
R.~S.~Chivukula, E.~H.~Simmons, H.~J.~He, M.~Kurachi and M.~Tanabashi,
{\it Deconstructed Higgsless models with one-site delocalization},
Phys.\ Rev.\ D {\bf 71}, 115001 (2005)
[arXiv:hep-ph/0502162].

\bibitem{Casalbuoni:2005rs}
R.~Casalbuoni, S.~De Curtis, D.~Dolce and D.~Dominici,
{\it Playing with fermion couplings in Higgsless models},
Phys.\ Rev.\ D {\bf 71}, 075015 (2005)
[arXiv:hep-ph/0502209].

\bibitem{SekharChivukula:2005xm}
R.~Sekhar Chivukula, E.~H.~Simmons, H.~J.~He, M.~Kurachi and M.~Tanabashi,
{\it Ideal fermion delocalization in Higgsless models},
Phys.\ Rev.\ D {\bf 72}, 015008 (2005)
[arXiv:hep-ph/0504114].

\bibitem{Son:2003et}
D.~T.~Son and M.~A.~Stephanov,
{\it QCD and dimensional deconstruction},
Phys.\ Rev.\ D {\bf 69}, 065020 (2004)
[arXiv:hep-ph/0304182].

\bibitem{Chivukula:2004kg}
R.~S.~Chivukula, M.~Kurachi and M.~Tanabashi,
{\it Generalized Weinberg sum rules in deconstructed QCD},
JHEP {\bf 0406}, 004 (2004)
[arXiv:hep-ph/0403112].

\bibitem{Sakai:2004cn}
T.~Sakai and S.~Sugimoto,
``Low energy hadron physics in holographic QCD,''
Prog.\ Theor.\ Phys.\  {\bf 113} (2005) 843
[arXiv:hep-th/0412141].

\bibitem{Erlich:2005qh}
J.~Erlich, E.~Katz, D.~T.~Son and M.~A.~Stephanov,
``QCD and a holographic model of hadrons,''
arXiv:hep-ph/0501128.

\bibitem{DaRold:2005zs}
L.~Da Rold and A.~Pomarol,
``Chiral symmetry breaking from five dimensional spaces,''
Nucl.\ Phys.\ B {\bf 721} (2005) 79
[arXiv:hep-ph/0501218].

\bibitem{Hirn:2005nr}
J.~Hirn and V.~Sanz,
``Interpolating between low and high energy QCD via a 5D Yang-Mills
model,''
arXiv:hep-ph/0507049.

\bibitem{Sakai:2005yt}
T.~Sakai and S.~Sugimoto,
``More on a holographic dual of QCD,''
arXiv:hep-th/0507073.


\bibitem{Hagiwara:1986vm}
  K.~Hagiwara, R.~D.~Peccei, D.~Zeppenfeld and K.~Hikasa,
  {\it Probing The Weak Boson Sector In E+ E- $\to$ W+ W-},
  Nucl.\ Phys.\ B {\bf 282}, 253 (1987).


\bibitem{Appelquist:1980vg}
T.~Appelquist and C.~W.~Bernard,
{\it Strongly Interacting Higgs Bosons},
Phys.\ Rev.\ D {\bf 22}, 200 (1980).

\bibitem{Longhitano:1980tm}
  A.~C.~Longhitano,
  {\it Low-Energy Impact Of A Heavy Higgs Boson Sector},
  Nucl.\ Phys.\ B {\bf 188}, 118 (1981).
  
\bibitem{Longhitano:1980iz}
  A.~C.~Longhitano,
  {\it Heavy Higgs Bosons In The Weinberg-Salam Model},
  Phys.\ Rev.\ D {\bf 22}, 1166 (1980).

\bibitem{Appelquist:1980ix}
T.~Appelquist,
{\it Broken Gauge Theories And Effective Lagrangians},
Print-80-0832 (YALE)
{\it Based on lectures presented at the 21st Scottish Universities Summer School in Physics, St. Andrews, Scotland, Aug 10-30, 1980}

\bibitem{Appelquist:1993ka}
  T.~Appelquist and G.~H.~Wu,
  {\it The Electroweak chiral Lagrangian and new precision measurements},
  Phys.\ Rev.\ D {\bf 48}, 3235 (1993)
  [arXiv:hep-ph/9304240].

\bibitem{Holdom:1990xq}
B.~Holdom,
{\it Corrections To Trilinear Gauge Vertices And E+ E- $\to$ W+ W- In Technicolor
Theories},
Phys.\ Lett.\ B {\bf 258}, 156 (1991).

\bibitem{Falk:1991cm}
A.~F.~Falk, M.~E.~Luke and E.~H.~Simmons,
{\it Chiral lagrangians and precision measurements of triple gauge boson vertices
at hadron colliders},
Nucl.\ Phys.\ B {\bf 365}, 523 (1991).

\bibitem{Gasser:1983yg}
  J.~Gasser and H.~Leutwyler,
  {\it Chiral Perturbation Theory To One Loop},
  Annals Phys.\  {\bf 158}, 142 (1984).
  
\bibitem{Georgi:2005dm}
  H.~Georgi,
  ``Chiral fermion delocalization in deconstructed Higgsless theories,''
  arXiv:hep-ph/0508014.

\bibitem{Birkedal:2004au}
A.~Birkedal, K.~Matchev and M.~Perelstein,
{\it Collider phenomenology of the Higgsless models},
Phys.\ Rev.\ Lett.\  {\bf 94}, 191803 (2005)
[arXiv:hep-ph/0412278].
 
 \bibitem{idealmodel}
  R.~S.~Chivukula, E.~H.~Simmons, H.~J.~He, M.~Kurachi and M.~Tanabashi,
  {\it Ideal Fermion Delocalization in Five Dimensional Gauge Theories},
  in preparation.
 
  \bibitem{LEPEWWG}
  The LEP Collaborations ALEPH, DELPHI, L3, OPAL and the LEP TGC Working Group.
  LEPEWWG/TC/2005-01; June 8, 2005.
  
\bibitem{Dobbs:2005ev}
  M.~Dobbs,
  {\it Prospects for Probing Triple Gauge-boson Couplings at the LHC},
  AIP Conf.\ Proc.\  {\bf 753}, 181 (2005)
  [arXiv:hep-ph/0506174].

\bibitem{Eboli:2004gc}
  O.~J.~P.~Eboli and M.~C.~Gonzalez-Garcia,
  {\it Probing trilinear gauge boson interactions via single electroweak gauge
  boson production at the LHC},
  Phys.\ Rev.\ D {\bf 70}, 074011 (2004)
  [arXiv:hep-ph/0405269].

\bibitem{Menges:2001gg}
  W.~Menges,
  {\it A study of charged current triple gauge couplings at TESLA},
LC-PHSM-2001-022

\bibitem{Boos:1999kj}
  E.~Boos, H.~J.~He, W.~Kilian, A.~Pukhov, C.~P.~Yuan and P.~M.~Zerwas,
  Phys.\ Rev.\ D {\bf 61}, 077901 (2000)
  [arXiv:hep-ph/9908409].
  
\bibitem{PDG}  S. Eidelman et al. (Particle Data Group), Physics Letters B592, 1 (2004) and 2005 partial updates for edition 2006 (URL:  http://pdg.lbl.gov). 

\bibitem{Godfrey:2000hc}
  S.~Godfrey, P.~Kalyniak, B.~Kamal and A.~Leike,
  {\it Discovery and identification of extra gauge bosons in  e+ e- $\to$ nu
  anti-nu gamma},
  Phys.\ Rev.\ D {\bf 61}, 113009 (2000)
  [arXiv:hep-ph/0001074].


\bibitem{Godfrey:2000pw}
  S.~Godfrey, P.~Kalyniak, B.~Kamal, M.~A.~Doncheski and A.~Leike,
  {\it Discovery and identification of W' bosons in e gamma $\to$ nu q + X},
  Phys.\ Rev.\ D {\bf 63}, 053005 (2001)
  [arXiv:hep-ph/0008157].

  
 \bibitem{Chivukula:inprep} 
R.~S. Chivukula, E.~H. Simmons, H.-J. He, M.~Kurachi, and M.~Tanabashi,
in preparation.

\bibitem{Csaki:2003sh}
C.~Csaki, C.~Grojean, J.~Hubisz, Y.~Shirman and J.~Terning,
``Fermions on an interval: Quark and lepton masses without a Higgs,''
Phys. Rev. D {\bf 70} (2004) 015012
[arXiv:hep-ph/0310355].

  
  
\end{thebibliography}

\end{document}